\documentclass[showpacs,aps,graphicx,twocolumn]{revtex4}
\usepackage{amsmath}
\usepackage{mathrsfs}
\usepackage{amsfonts}
\usepackage{amssymb}
\usepackage{graphicx}
\usepackage{subfigure}
\usepackage{eufrak}
\usepackage{multirow}
\usepackage{longtable}
\usepackage{float}
\usepackage{bm}
\usepackage{xcolor}
\usepackage[colorlinks,linkcolor=blue,anchorcolor=blue,citecolor=blue,urlcolor=blue]{hyperref}
\usepackage{soul,color,xcolor}

\begin{document}

\title{Optimal synthesis of general multi-qutrit quantum computation}

\author{Gui-Long Jiang$^{1}$, Wen-Qiang Liu$^{2}$ and Hai-Rui Wei$^{1}$}
\email[]{hrwei@ustb.edu.cn}
\address{$^{1}$School of Mathematics and Physics, University of Science and Technology Beijing, Beijing 100083, China\\
         $^{2}$Center for Quantum Technology Research and Key Laboratory of Advanced Optoelectronic Quantum Architecture and Measurements (MOE), School of Physics, Beijing Institute of Technology, Beijing 100081, China\\}

\begin{abstract}
Quantum circuits of a general quantum gate acting on multiple $d$-level quantum systems play a prominent role in multi-valued quantum computation.
We first propose a new recursive Cartan decomposition of semi-simple unitary Lie group $U(3^n)$ (arbitrary $n$-qutrit gate). Note that the decomposition completely decomposes an $n$-qutrit gate into local and
non-local operations.
We design an explicit quantum circuit for implementing arbitrary two-qutrit gates, and the cost of our construction is 21 generalized controlled $X$ (GCX) and controlled increment (CINC) gates less than the earlier best result of 26 GGXs.
Moreover, we extend the program to the $n$-qutrit system, and the quantum circuit of generic $n$-qutrit gates contained  $\frac{41}{96}\cdot3^{2n}-4\cdot3^{n-1}-(\frac{n^2}{2}+\frac{n}{4}-\frac{29}{32})$ GGXs and CINCs is presented. Such asymptotically optimal structure is the best known result so far.
\end{abstract}


\maketitle

\section{Introduction}\label{sec1}

In recent decades, quantum computing has been extensively researched and has gradually formed a complete theoretical system \cite{1}.
Compared to established computational theory, quantum computers are qualitatively stronger \cite{2,3,4} and capable of solving classical computational puzzles, e.g., Grover's algorithm \cite{Grover1,Grover2}, Shor's algorithm \cite{Shor}, and Deutsch-Jozsa algorithm \cite{Deutsch1992}, more efficiently due to the powerful features of quantum mechanics.
It is well known that a $d$-level quantum system, so-called qudit (qubit when $d=2$; qutrit when $d=3$), and its state space can be described by a complex $d$-dimensional Hilbert space $\mathcal{H}_{d}$ with the basis $\{|0\rangle,|1\rangle,\cdots,|d-1\rangle\}$.
Higher-dimensional Hilbert space exhibits unique advantages such as enlarging channel capacities \cite{capacity}, improving resistance to noise \cite{noise,Wang2022}, supporting fault tolerance \cite{tolerance}, enhancing robustness to quantum cloning \cite{clone}, reducing the experimental requirements and  quantum resource overheads  \cite{resource}, completing some tasks which are impossible to achieve with standard qubit-based technologies \cite{task1,task2,task3}, and so on.
Nowadays, tremendous progress both in theory and experiment has been made in qudit-based quantum information processing, e.g., $d$-dimensional quantum gate \cite{high-comput1,high-comput2}, Grover's algorithm \cite{high-algorithm}, entangled state \cite{high-entagled1,high-entagled2,high-entagled3}, quantum teleportation \cite{high-teleportation1,high-teleportation2,high-teleportation3}, quantum key distribution \cite{QKD1,QKD2}, and quantum tomography \cite{Tomography}.

The quantum circuit model is the most popular paradigm for implementing complex quantum computing \cite{circuit1,circuit2}.
An efficient synthesis of quantum circuits should be as low-cost and shallow as possible because the growth of qubits increases the length and time scales of quantum gates, leads to more complex procedures, and makes the quantum system further susceptible to their environment \cite{circuit3,circuit4}.
The controlled-NOT (CNOT) gate is the most popular universal gate for qubit-based quantum computing \cite{Barenco1995}.
Nowadays, using CNOT gates as elementary gates of synthesis, the optimal quantum circuit for arbitrary two-qubit gates has been completed \cite{small1,small2}, and considerable progress has been made on the physical implementation of CNOT \cite{CNOT1,CNOT2,CNOT3}.
To simplify multi-qubit quantum computation, several synthesis programs are proposed using QR decomposition \cite{Barenco1995,Vartiainen2004}, Cosine-Sine decomposition (CSD) \cite{Mottonen2004}, quantum Shannon decomposition (QSD) \cite{Shende2006}, and some specific Cartan decompositions \cite{Khaneja2001,Bullock2004,Alessandro2007}.
However, the above results are much higher than the lower bound of $\frac{1}{4}(4^n-3n-1)]$ CNOT gates \cite{small2}.
Lanyon \emph{et al}. \cite{task1,task2} showed that the cost of the qubit-based quantum circuit may be further reduced using higher-dimensional Hilbert spaces.

Qudit-based quantum computation is in the infant.
Earlier in 2003, universal gates GXOR \cite{GXOR}, SUM, and double SUM \cite{sum} for $d$-level system were proposed, but the quantum circuits were not constructed in terms of these gates.
Bullock \emph{et al}. \cite{Bullock2005} in 2005 proposed an asymptotically optimal quantum circuit for implementing arbitrary $n$-qudit gates where $\Theta(d^{2n})$ two-qudit gates are required.
They later proved that the controlled increment (CINC) gate is universal and presented another asymptotically optimal quantum synthesis for $d$-level quantum computation using CINC as the elementary gate \cite{Brennen2005}.
In addition, based on CSD, Nakajima \emph{et al.} \cite{Nakajima2009} suggested a synthesis algorithm for any operations on $n$-qudits, which reduces the 156 CINCs required in Ref. \cite{Brennen2005} to 36 in the case of two-qutrit.
A universal generalized controlled $X$ (GCX) gate was proposed by Di \emph{ et al}. \cite{Di2013} in 2013, and the GCX-based simplified synthesis of a generic $n$-qudit gate was presented with extended QSD.
Later in 2015, they \cite{Di2015} refined CSD to further reduce the cost (i.e., the number of GCX gates) of synthetic circuits for an arbitrary $n$-qudit gate.
Nevertheless, this is still far from the theoretical lower bound of $[d^{2n}-n(d^2-1)-1]/[4(d-1)]$ GCXs \cite{Di2015}.
In particular, for two-qutrit ($d=3$, $n=2$), the cost of optimal synthesis is currently 26 GCXs \cite{Di2015}, while the theoretical lower bound is only 8.
Additionally, so far the efficient synthesis for a general $n$-qutrit gate via Cartan decomposition remains an open question.

In this paper, we give an explicit program to synthesize arbitrary $n$-qutrit computation with the shortest possible cost.
We first propose a recursive Cartan decomposition of the unitary operator acting on $n$-qutrit, and such a new recursive algorithm first completely decomposes generic $n$-qutrit gates into a product of simple local and non-local components.
Based on the decomposition shown in Fig. \ref{Fig.9}, we then present an explicit quantum circuit for generic $n$-qutrit gates in terms of single-qutrit type rotations and two-qutrit GCX gates.
Note that $\frac{47}{96}\cdot3^{2n}-4\cdot3^{n-1}-(\frac{n^2}{2}+\frac{3n}{4}-\frac{27}{32})$ GCX gates contained in our construction are minimal compared to the previous ones.
If restricted to universal CINC and GCX gates, the cost can be further reduced to $\frac{41}{96}\cdot3^{2n}-4\cdot3^{n-1}-(\frac{n^2}{2}+\frac{n}{4}-\frac{29}{32})$ elementary two-qutrit gates, and only 21 CINC and GCX gates are required in the worst to simulate an arbitrary two-qutrit gate, while the previous optimum one required 26 \cite{Di2015}.

The paper is organized as follows: In Sec. \ref{sec2}, we briefly review the Cartan decomposition, the elementary-gate library, and propose two Theorems.
In Sec. \ref{sec3}, the Cartan decomposition of an arbitrary two-qutrit gate is proposed and a synthesis of generic two-qutrit gates is designed.
In Sec. \ref{sec4}, this program is promoted to synthesize $n$-qutrit gates, and a rigorous proof for the Cartan decomposition of $n$-qutrit gates is presented.
Finally, the conclusion is provided in Sec. \ref{sec5}.

\section{Cartan decomposition and three-valued elementary gate library}\label{sec2}

First, let us briefly review the Cartan decomposition of Lie groups and three-valued elementary gate library containing the single-qutrit gate $R_\varphi^{ij}(\theta)$ and two-qutrit GCX gate, and Theorem 1 and Theorem 2 are proposed.

\subsection{Cartan decomposition of Lie group}\label{sec2.1}

Assume that $G$ is a compact semi-simple Lie group and $\mathfrak{g}$ is its associated real semi-simple Lie algebra.
Given a direct sum decomposition of $\mathfrak{g}$ \cite{cartan}
\begin{eqnarray}\label{eq1}
\mathfrak{g}=\mathfrak{l} \oplus \mathfrak{p},
\end{eqnarray}
if the Lie subalgebra $\mathfrak{l}$ and its complement space $\mathfrak{p}$ satisfy the following commutation relations:
\begin{eqnarray}\label{eq2}
[\mathfrak{l},\mathfrak{l}]\subseteq \mathfrak{l},\;\;
[\mathfrak{l},\mathfrak{p}]\subseteq \mathfrak{p},\;\;
[\mathfrak{p},\mathfrak{p}]\subseteq \mathfrak{l},
\end{eqnarray}
then Eq. (\ref{eq1}) is called Cartan decomposition of $\mathfrak{g}$.
If $\mathfrak{a}$ is a maximally Abelian subalgebra contained in $\mathfrak{p}$, then $\mathfrak{a}$ is called a Cartan subalgebra of the pair $(\mathfrak{g},\mathfrak{l})$.

Based on Eq. (\ref{eq1}) and the relation between the Lie group and the Lie algebra, any element $M\in G$ can be decomposed as
\begin{eqnarray}\label{eq3}
M=K_1\cdot A \cdot K_2,
\end{eqnarray}
where $K_1,K_2\in \mathrm{exp}(\mathfrak{l})$ and $A\in \mathrm{exp}(\mathfrak{a})$.
Eq. (\ref{eq3}) is called Cartan decomposition of the Lie group $G$.

\subsection{Synthesis of single-qutrit gate}\label{sec2.2}
Let $U(3^n)$ denote the unitary group of $3^n\times3^n$ unitary matrices and $\mathfrak{u}(3^n)$ be its associated Lie algebra.
It is known that the matrix of any single-qutrit gate is an element of $U(3)$.
The basis of $\mathfrak{u}(3)$, where $\mathfrak{u}(3)$ is a 9-dimensional linear space of $3\times3$ skew-Hermitian matrices, can be given by \cite{Di2008}
\begin{eqnarray}\label{eq4}
\begin{aligned}
\mathfrak{u}(3)=\textrm{span}
\{&\textrm{i}\sigma_x^{01},\textrm{i}\sigma_x^{02},\textrm{i}\sigma_x^{12},
\textrm{i}\sigma_y^{01},\textrm{i}\sigma_y^{02},\\&\textrm{i}\sigma_y^{12},
\textrm{i}\sigma_z^{01},\textrm{i}\sigma_z^{02},\textrm{i}I_3\},
\end{aligned}
\end{eqnarray}
where
\begin{eqnarray}\label{eq5}
\begin{aligned}
\sigma_{x}^{01}&=\left(
  \begin{array}{ccc}
    0 & 1 & 0\\
    1 & 0 & 0\\
    0 & 0 & 0\\
  \end{array}
\right),\;\;\;\;
\sigma_{x}^{02}=\left(
  \begin{array}{ccc}
    0 & 0 & 1\\
    0 & 0 & 0\\
    1 & 0 & 0\\
  \end{array}
\right),\\
\sigma_{x}^{12}&=\left(
  \begin{array}{ccc}
    0 & 0 & 0\\
    0 & 0 & 1\\
    0 & 1 & 0\\
  \end{array}
\right),\;\;\;\;
\sigma_{y}^{01}=\left(
  \begin{array}{cccc}
    0            & -\mathrm{i} & 0\\
    \mathrm{i}   & 0           & 0\\
    0            & 0            & 0\\
  \end{array}
\right),\\
\sigma_{y}^{02}&=\left(
  \begin{array}{cccc}
    0 & 0 & -\mathrm{i}\\
    0 & 0 & 0\\
    \mathrm{i} & 0 & 0\\
  \end{array}
\right),\;\;
\sigma_{y}^{12}=\left(
  \begin{array}{ccc}
    0 & 0          & 0\\
    0 & 0          & -\mathrm{i}\\
    0 & \mathrm{i} & 0\\
  \end{array}
\right),\\
\sigma_{z}^{01}&=\left(
  \begin{array}{cccc}
    1 & 0 & 0\\
    0 & -1 & 0\\
    0 & 0 & 0\\
  \end{array}
\right),\;
\sigma_{z}^{02}=\left(
  \begin{array}{cccc}
    1 & 0 & 0\\
    0 & 0 & 0\\
    0 & 0 & -1\\
  \end{array}
\right),\\
I_3&=\left(
  \begin{array}{cccc}
    1 & 0 & 0\\
    0 & 1 & 0\\
    0 & 0 & 1\\
  \end{array}
\right).
\end{aligned}
\end{eqnarray}
Here and afterward, span$\{A\}$ denotes the space given by all linear combinations of elements in $A$ with real coefficients.

Imitating single-qubit rotation gates, we define a single-qutrit $\varphi$-axes type rotation gate as
\begin{eqnarray}\label{eq6}
R_\varphi^{ij}(\theta)=\textrm{exp}(-\textrm{i}\frac{\theta}{2}\sigma_\varphi^{ij})
\end{eqnarray}
where $\varphi \in \{x,y,z\}$, $i,j\in\{0,1,2\}$.
Based on the $\mathrm{AIII}$-type Cartan decomposition of $U(3)$, an arbitrary single-qutrit gate can be simulated by certain $R_z^{ij}(\theta)$ ($R_x^{ij}(\theta)$) gates and $R_y^{i'j'}(\theta)$ gates \cite{Di2013}. In addition, we present the transformation relation between $R_x^{ij}(\theta)$ and $R_z^{ij}(\theta)$.

\textbf{Theorem 1.} $R_y^{ij}(-\frac{\pi}{2}) \cdot R_x^{ij}(\theta) \cdot R_y^{ij}(\frac{\pi}{2}) =R_z^{ij}(\theta)$.

\textbf{Proof 1.} This can be obtained directly from matrix multiplication.$\hfill\blacksquare$

\subsection{Two-qutrit universal gate}\label{sec2.3}

Let $\{|0\rangle,|1\rangle,|2\rangle\}$ be the base state of a single-qutrit system. The generalized controlled $X$ gate, $\textrm{GCX}_l^k(m\rightarrow X^{ij})$, can be recognized as a two-qutrit universal gate \cite{Di2013}.
Here $\textrm{GCX}_l^k(m\rightarrow X^{ij})$ is to perform an $X^{ij}$ operation on the target qutrit (the $k$th qutrit) if and only if the state of the control qutrit (the $l$th qutrit) is $|m\rangle$ $(m \in\{0,1,2\})$.
In the $\{|0\rangle, |1\rangle, |2\rangle\}$ basis, $X^{01}$, $X^{02}$, and $X^{12}$ are given by
\begin{eqnarray}\label{eq7}
\begin{aligned}
X^{01}&=\left(
  \begin{array}{ccc}
    0 & 1 & 0\\
    1 & 0 & 0\\
    0 & 0 & 1\\
  \end{array}
\right),\\
X^{02}&=\left(
  \begin{array}{ccc}
    0 & 0 & 1\\
    0 & 1 & 0\\
    1 & 0 & 0\\
  \end{array}
\right),\\
X^{12}&=\left(
  \begin{array}{ccc}
    1 & 0 & 0\\
    0 & 0 & 1\\
    0 & 1 & 0\\
  \end{array}
\right).
\end{aligned}
\end{eqnarray}
Some two adjacent GCX gates can be simplified as a GCX gate and an $X^{ij}$ gate.

\textbf{Theorem 2.} For an $n$-qutrit system, if $m, m', m''\in \{0,1,2\}$ are not equal to each other, there is
\begin{eqnarray}\label{eq8}
\begin{aligned}
&\textrm{GCX}_l^k(m\rightarrow X^{ij})\cdot \textrm{GCX}_l^k(m'\rightarrow X^{ij})\\&=\textrm{GCX}_l^k(m''\rightarrow X^{ij})\cdot I_{3^{k-1}}\otimes X^{ij}\otimes I_{3^{n-k}}
\\&=I_{3^{k-1}}\otimes X^{ij}\otimes I_{3^{n-k}} \cdot \textrm{GCX}_l^k(m''\rightarrow X^{ij}).
\end{aligned}
\end{eqnarray}
Here $I_d$ is the $d\times d$ identity matrix.

\textbf{Proof 2.} The conclusion follows directly from matrix multiplication.$\hfill\blacksquare$

\begin{figure} [htbp]
  \centering
  \subfigure[]{
  \includegraphics[width=5.8cm]{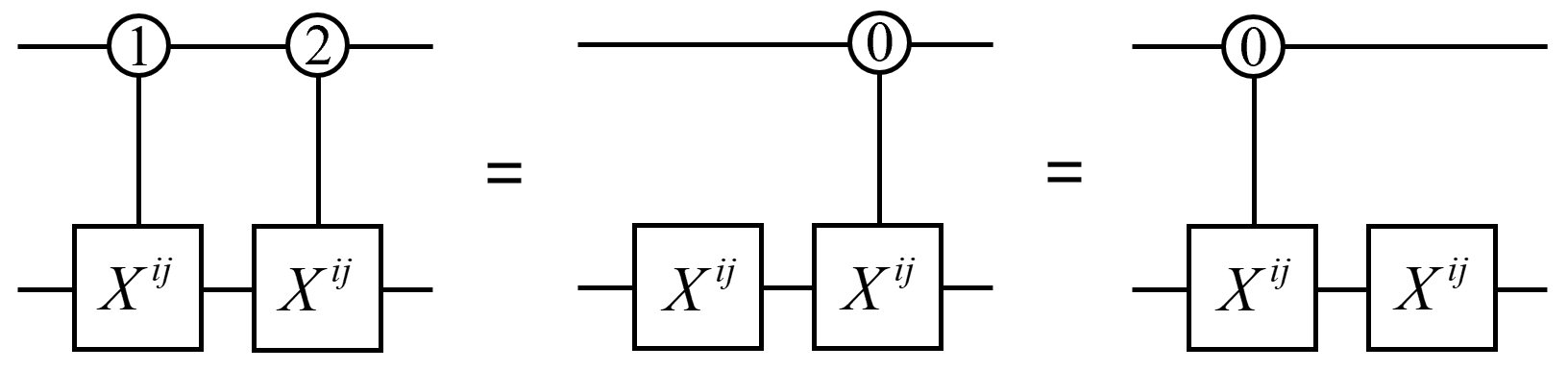}
 \label{1a} }
  \subfigure[]{
  \includegraphics[width=3cm]{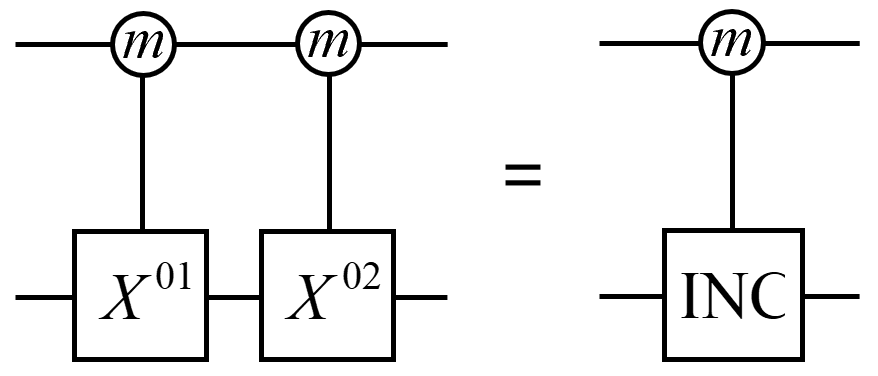}
  \label{1b}}
  \caption{(a) The quantum circuit of Theorem 2 with $n=2$, $l=1$, $k=2$, and $mm'm''=210$. (b) The quantum circuit of Eq. (\ref{eq10}) with $l=1$ and $k=2$.}
  \label{Fig.1}
\end{figure}

In particular, with a two-qutrit system, when $l=1$ and $k=2$, Theorem 2 can be illustrated by the quantum circuit shown in Fig. \ref{1a}.

Aside from GCX gates, the controlled increment gate, $\textrm{CINC}_l^k(m\rightarrow \textrm{INC})$, is another common two-qutrit universal gate \cite{Di2013}. Here $\textrm{CINC}_l^k(m\rightarrow \textrm{INC})$ is defined as implementing the INC operation on the target qutrit (the $k$th qutrit) if and only if the state of the control qutrit (the $l$th qutrit) is $|m\rangle$, where
\begin{eqnarray}\label{eq9}
\begin{aligned}
\textrm{INC}=X^{02}\cdot X^{01}&=\left(
  \begin{array}{ccc}
    0 & 0 & 1\\
    1 & 0 & 0\\
    0 & 1 & 0\\
  \end{array}
\right).
\end{aligned}
\end{eqnarray}
By Eq. (\ref{eq9}), it is easy to obtain that
\begin{eqnarray}\label{eq10}
\begin{aligned}
&\textrm{GCX}_l^k(m\rightarrow X^{02})\cdot \textrm{GCX}_l^k(m\rightarrow X^{01})\\&=\textrm{CINC}_l^k(m\rightarrow \textrm{INC}).
\end{aligned}
\end{eqnarray}
When $l=1$ and $k=2$, the quantum circuit represented by the above equation is shown in Fig. \ref{1b}.

\section{Cartan decomposition of arbitrary two-qutrit gates}\label{sec3}

\subsection{Cartan decomposition of Lie group $\mathbf{\emph{U}(9)}$}\label{sec3.1}

We provide a successive cartan decomposition of $U(9)$, and refer to Sec. \ref{sec4.1} for rigorous proof.
Without loss of generality, the Lie algebra $\mathfrak{u}(9)$ can be written as
\begin{eqnarray}\label{eq11}
\begin{aligned}
\mathfrak{u}(9)=\textrm{span}
\{&\sigma_x^{01}\otimes A,\sigma_x^{02}\otimes A,\sigma_x^{12}\otimes A,\sigma_y^{01}\otimes A,\\&\sigma_y^{02}\otimes A,\sigma_y^{12}\otimes A,\sigma_z^{01}\otimes A,\sigma_z^{02}\otimes A,\\&I_3\otimes A|A\in \mathfrak{u}(3)\}.
\end{aligned}
\end{eqnarray}

\textbf{Decomposition of} \bm{$\mathfrak{u}(9)$}. We define
\begin{eqnarray}\label{eq12}
\begin{aligned}
\mathfrak{l}(9)=\textrm{span}
\{&\sigma_x^{12}\otimes A,\sigma_y^{12}\otimes A,\sigma_z^{01}\otimes A,\\&\sigma_z^{02}\otimes A,I_3\otimes A|A\in \mathfrak{u}(3)\},\\
\mathfrak{p}(9)=\textrm{span}
\{&\sigma_x^{01}\otimes A,\sigma_x^{02}\otimes A,\sigma_y^{01}\otimes A,\\&\sigma_y^{02}\otimes A|A\in \mathfrak{u}(3)\}.
\end{aligned}
\end{eqnarray}
Applying the formula
\begin{eqnarray}\label{eq13}
[A\otimes B,C\otimes D]=[A,C]\otimes(BD)+(CA)\otimes [B,D],
\end{eqnarray}
it is easily verified that $\mathfrak{u}(9)=\mathfrak{l}(9)\oplus \mathfrak{p}(9)$ is the Cartan decomposition of $\mathfrak{u}(9)$ and the corresponding Cartan subalgebra $\mathfrak{a}(9)$ can be given by
\begin{eqnarray}\label{eq14}
\mathfrak{a}(9)=\textrm{span}
\{\textrm{i}\sigma_x^{01}\otimes \sigma_z^{01},\textrm{i}\sigma_x^{01}\otimes \sigma_z^{02},\textrm{i}\sigma_x^{01}\otimes I_3\}.
\end{eqnarray}

\textbf{Decomposition of} \bm{$\mathfrak{l}(9)$}. Note that Lie subalgebra $\mathfrak{l}(9)$ is isomorphic to $\mathfrak{u}(3)\oplus \mathfrak{u}(6)$, and it can be further decomposed as $\mathfrak{l}(9)=\mathfrak{l}_1(9)\oplus \mathfrak{p}_1(9)$ with
\begin{eqnarray}\label{eq15}
\begin{aligned}
&\mathfrak{l}_{1}(9)=\textrm{span}\{\sigma_z^{01}\otimes A,\sigma_z^{02}\otimes A,I_3\otimes A|A\in \mathfrak{u}(3)\},\\
&\mathfrak{p}_{1}(9)=\textrm{span}\{\sigma_x^{12}\otimes A,\sigma_y^{12}\otimes A,|A\in \mathfrak{u}(3)\}.
\end{aligned}
\end{eqnarray}
The corresponding Cartan subalgebra can be given by
\begin{eqnarray}\label{eq16}
\mathfrak{a}_1(9)=\textrm{span}\{\textrm{i}\sigma_x^{12}\otimes \sigma_z^{01},\textrm{i}\sigma_x^{12}\otimes \sigma_z^{02},\textrm{i}\sigma_x^{12}\otimes I_3\}.
\end{eqnarray}

\textbf{Decomposition of} \bm{$\mathfrak{l}_1(9)$}. Lie subalgebra $\mathfrak{l}_1(9)$ is isomorphic to $\mathfrak{u}(3)\oplus \mathfrak{u}(3) \oplus \mathfrak{u}(3)$. In order to faithfully separate the local contents from the nonlocal contained in $\mathfrak{l}_1(9)$, we rewrite the diagonal base elements of $\mathfrak{u}(3)$ as
\begin{eqnarray}\label{eq17}
\begin{aligned}
\textrm{span}\{I_3,\sigma_z^{01},\sigma_z^{02}\}
=&\textrm{span}\{I_3,D,\sigma_z^{12}\}\\
=&\textrm{span}\{I_3,\overline{D},\sigma_z^{01}\}.
\end{aligned}
\end{eqnarray}
Here
\begin{eqnarray}\label{eq18}
\begin{aligned}
\sigma_{z}^{12}=&\left(
  \begin{array}{ccc}
    0 & 0 & 0\\
    0 & 1 & 0\\
    0 & 0 & -1\\
  \end{array}
\right)=\sigma_{z}^{02}-\sigma_{z}^{01},\\
D=&\left(
  \begin{array}{ccc}
    1 & 0 & 0\\
    0 & -1 & 0\\
    0 & 0 & -1\\
  \end{array}
\right)=\frac{-I_3+2\sigma_{z}^{01}+2\sigma_{z}^{02}}{3},\\
\overline{D}=&\left(
  \begin{array}{ccc}
    -1 & 0 & 0\\
    0 & -1 & 0\\
    0 & 0 & 1\\
  \end{array}
\right)=\frac{-I_3+2\sigma_{z}^{01}-4\sigma_{z}^{02}}{3}.
\end{aligned}
\end{eqnarray}

Then $\mathfrak{l}_1(9)$ can be written as
\begin{eqnarray}\label{eq19}
\mathfrak{l}_1(9)=\textrm{span}\{I_3\otimes A,D\otimes A,\sigma_z^{12}\otimes A|A\in \mathfrak{u}(3)\}.
\end{eqnarray}
We have $\mathfrak{l}_1(9)=\mathfrak{l}_2(9)\oplus \mathfrak{p}_2(9)$ and the Cartan subalgebra $\mathfrak{a}_2(9)$ of the pair $(\mathfrak{l}_1(9),\mathfrak{l}_2(9))$, where
\begin{eqnarray}\label{eq20}
\begin{aligned}
&\mathfrak{l}_2(9)=\textrm{span}\{I_3\otimes A,D\otimes A|A\in \mathfrak{u}(3)\},\\
&\mathfrak{p}_2(9)=\textrm{span}\{\sigma_z^{12}\otimes A|A\in \mathfrak{u}(3)\},\\
&\mathfrak{a}_2(9)=\textrm{span}
\{\textrm{i}\sigma_z^{12}\otimes I_3,\textrm{i}\sigma_z^{12}\otimes D,\textrm{i}\sigma_z^{12}\otimes \sigma_z^{12}\}.
\end{aligned}
\end{eqnarray}

When $\textrm{span}\{I_3,\overline{D},\sigma_z^{01}\}$ is chosen,
$\mathfrak{l}_1(9)$ can be rewritten as
\begin{eqnarray}\label{eq21}
\mathfrak{l}_1(9)=\textrm{span}\{I_3\otimes A,\overline{D}\otimes A,\sigma_z^{01}\otimes A|A\in \mathfrak{u}(3)\},
\end{eqnarray}
and has the Cartan decomposition
\begin{eqnarray}\label{eq22}
\begin{aligned}
&\mathfrak{l}_1(9)=\overline{\mathfrak{l}_2}(9)\oplus \overline{\mathfrak{p}_2}(9),\\
&\overline{\mathfrak{l}_2}(9)=\textrm{span}\{I_3\otimes A,\overline{D}\otimes A|A\in \mathfrak{u}(3)\},\\
&\overline{\mathfrak{p}_2}(9)=\textrm{span}\{\sigma_z^{01}\otimes A|A\in \mathfrak{u}(3)\}.
\end{aligned}
\end{eqnarray}

\textbf{Decomposition of \bm{$\mathfrak{l}_2(9)$} and \bm{$\overline{\mathfrak{l}_2}(9)$}}. Lie algebra $\mathfrak{l}_2(9)$ ($\overline{\mathfrak{l}_2}(9)$) is isomorphic to $\mathfrak{u}(3)\oplus \mathfrak{u}(3)$. Then the local and nonlocal terms contained in $\mathfrak{l}_2(9)$ can be successfully divided as
\begin{eqnarray}\label{eq23}
\begin{aligned}
&\mathfrak{l}_2(9)=\mathfrak{l}_3(9)\oplus \mathfrak{p}_3(9),\\
&\mathfrak{l}_3(9)=\textrm{span}
\{I_3\otimes A|A\in \mathfrak{u}(3)\},\\
&\mathfrak{p}_3(9)=\textrm{span}
\{D  \otimes A|A\in \mathfrak{u}(3)\}.
\end{aligned}
\end{eqnarray}
The Cartan subalgebra $\mathfrak{a}_3(9)$ of the pair $(\mathfrak{l}_2(9),\mathfrak{l}_3(9))$ is given by
\begin{eqnarray}\label{eq24}
\mathfrak{a}_3(9)=\textrm{span}
\{\textrm{i}D\otimes I_3,\textrm{i}D\otimes D,\textrm{i}D\otimes \sigma_z^{12}\}.
\end{eqnarray}
Likewise, for $\overline{\mathfrak{l}_2}(9)$, we have
\begin{eqnarray}\label{eq25}
\begin{aligned}
&\overline{\mathfrak{l}_2}(9)=\mathfrak{l}_3(9)\oplus \overline{\mathfrak{p}_3}(9),\\
&\overline{\mathfrak{p}_3}(9)=\textrm{span}
\{\overline{D}\otimes A|A\in \mathfrak{u}(3)\},\\
&\overline{\mathfrak{a}_3}(9)=\textrm{span}
\{\textrm{i}\overline{D}\otimes I_3,\textrm{i}\overline{D}\otimes \overline{D},\textrm{i}\overline{D}\otimes \sigma_z^{01}\}.
\end{aligned}
\end{eqnarray}

\begin{figure} [htbp]
  \centering
  \subfigure[]{
  \includegraphics[width=5.5cm]{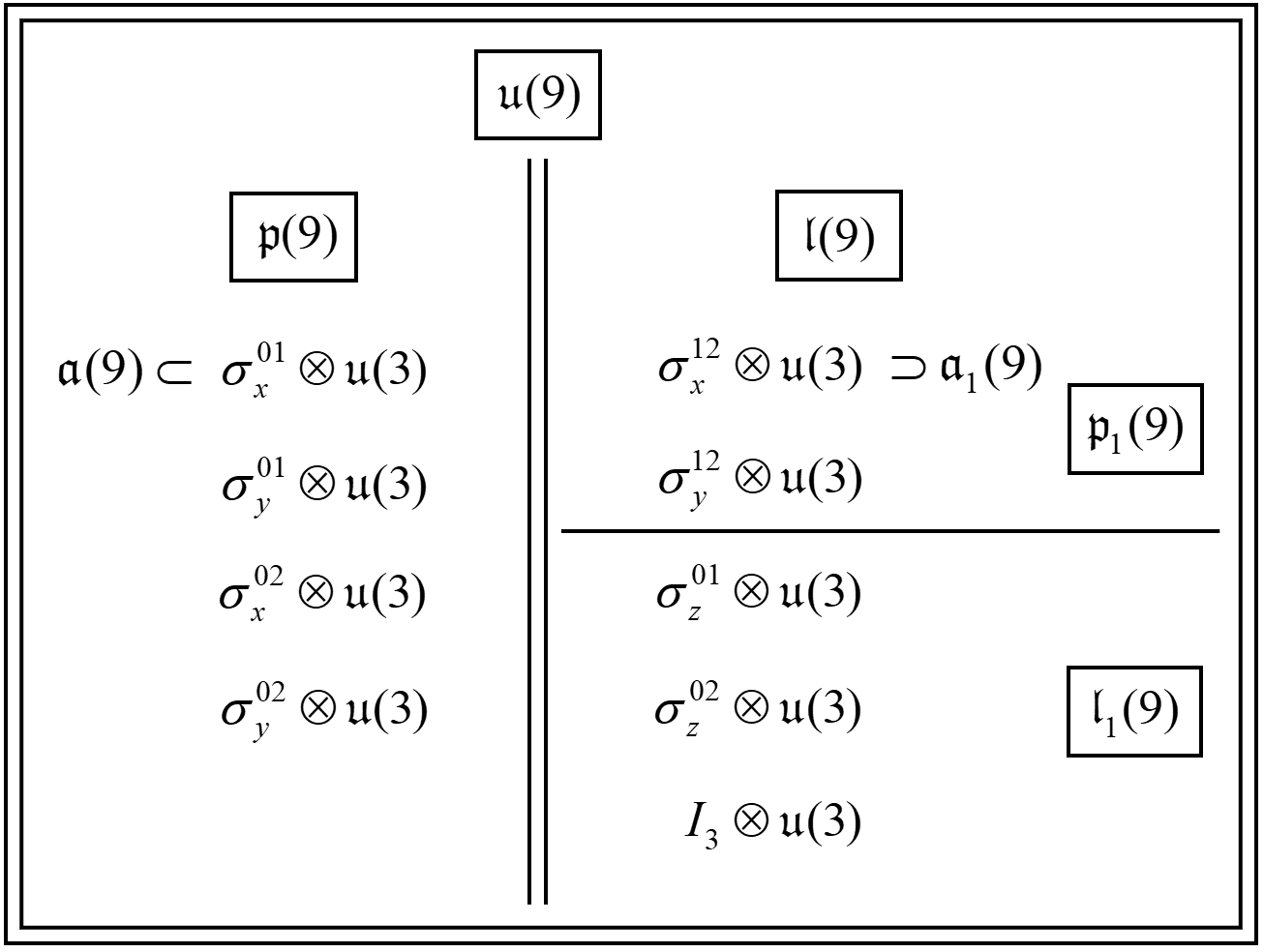}
 \label{2a} }
  \subfigure[]{
  \includegraphics[width=4.05cm]{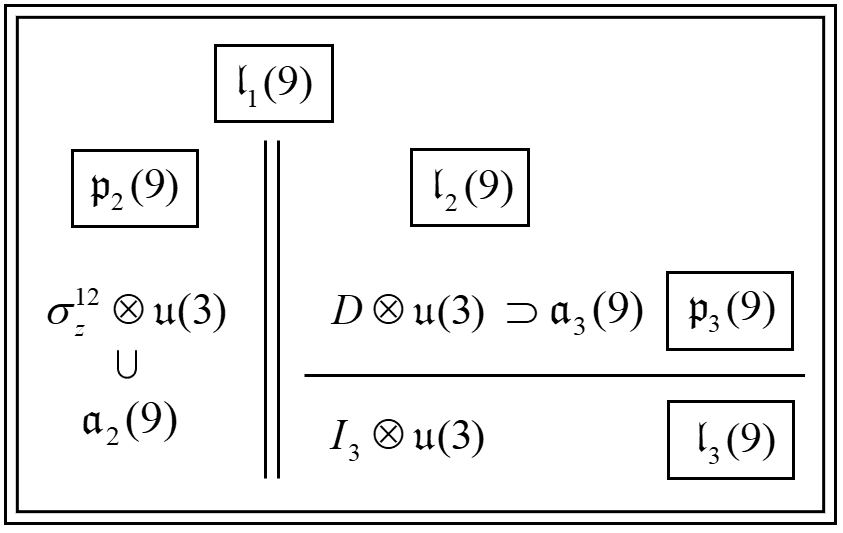}
  \label{2b}}
  \subfigure[]{
  \includegraphics[width=4.05cm]{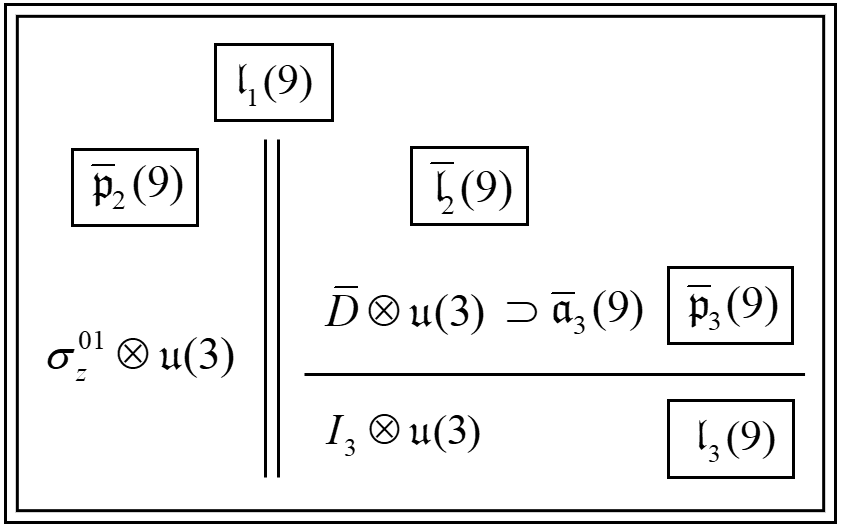}
  \label{2c}}
  \caption{(a) Cartan decomposition of $\mathfrak{u}(9)$ and $\mathfrak{l}(9)$.
  (b) Cartan decomposition of $\mathfrak{l}_1(9)$ and $\mathfrak{l}_2(9)$ with the diagonal basis $\{I_3,D,\sigma_z^{12}\}$. (c) Cartan decomposition of $\mathfrak{l}_1(9)$ and $\overline{\mathfrak{l}_2}(9)$ with the diagonal basis $\{I_3,\overline{D},\sigma_z^{01}\}$. Here $A\otimes\mathfrak{u}(3)$ denotes the set obtained by tensoring matrix $A$ with each element of $\mathfrak{u}(3)$, i.e., $\{A\otimes B|B\in\mathfrak{u}(3)\}$.}
  \label{Fig.2}
\end{figure}

Based on the above cascaded Cartan decomposition depicted by Fig. \ref{Fig.2} and the relation between Lie algebra and Lie group, the following Corollary can be obtained, for a proof of which see Appendix \ref{Appendix.A}.

\textbf{Corollary 1.} Any $M\in U(9)$ can be decomposed as
\begin{eqnarray}\label{eq26}
\begin{aligned}
M=&K_1 \cdot \overline{A}^{(3)}_1 \cdot K_2 \cdot A^{(1)}_1 \cdot K_3 \cdot A^{(3)}_1 \cdot K_4 \cdot A \cdot K_5\\ &\cdot \overline{A}^{(3)}_2 \cdot K_6 \cdot A^{(1)}_2 \cdot K_7 \cdot A^{(2)} \cdot K_8 \cdot A^{(3)}_2\cdot K_9.
\end{aligned}
\end{eqnarray}
Here $A\in \textrm{exp}[\mathfrak{a}(9)]$, $A^{(1)}_{1,2}\in \textrm{exp}[\mathfrak{a}_1(9)]$, $A^{(2)}\in \textrm{exp}[\mathfrak{a}_2(9)]$, $A^{(3)}_{1,2}\in \textrm{exp}[\mathfrak{a}_3(9)]$,
$\overline{A}^{(3)}_{1,2}\in \textrm{exp}[\overline{\mathfrak{a}_3}(9)]$, and $K_i\in \textrm{exp}[\mathfrak{l}_3(9)]$.
Note that $K_l\in U(3)$ as $\mathfrak{l}_3(9)$ is isomorphic to $\mathfrak{u}(3)$.

\subsection{Synthesis of arbitrary two-qutrit gates}\label{sec3.2}

\begin{figure} [htbp]
  \centering
  \includegraphics[width=8.8cm]{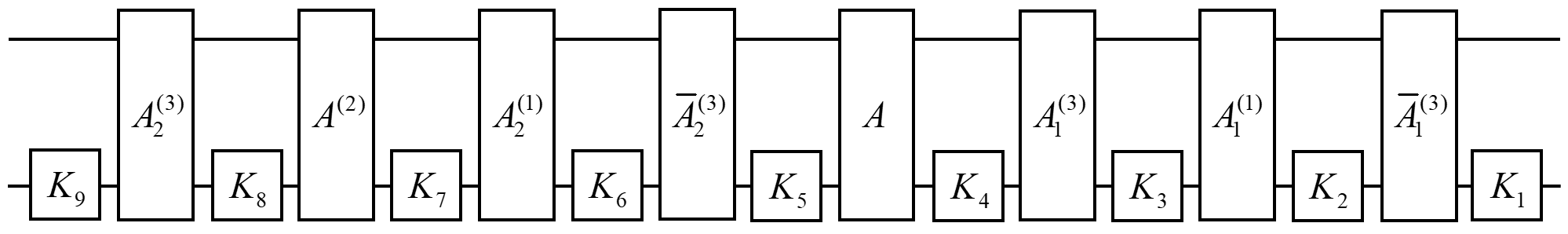}
  \caption{Decomposition of an arbitrary two-qutrit gate based on Eq. (\ref{eq26}).}
  \label{Fig.3}
\end{figure}

As shown in Fig. \ref{Fig.3}, two-qutrit nonlocal operations $A$, $A_i^{(1)}$, $A_j^{(2)}$, and $A_k^{(3)}$ play the key role in the construction of an arbitrary two-qutrit gate. In the following, we provide a program to synthesize them in detail.

\textbf{Synthesis of two-qutrit operation \bm{$A$}}.
Based on Eq. (\ref{eq14}), nonlocal operation $A$ can be expressed as
\begin{eqnarray}\label{eq27}
\begin{aligned}
A(\theta_1,\theta_2,\theta_3)=&\textrm{exp}[-\textrm{i}(\theta_1  \sigma_x^{01}\otimes\sigma_z^{01}+\theta_2\sigma_x^{01}\otimes\sigma_z^{02}\\&
+\theta_3\sigma_x^{01}\otimes I_3)].
\end{aligned}
\end{eqnarray}

By employing Theorem 1, we find that $A(\theta_1,\theta_2,\theta_3)$ is equivalent to $\widetilde{A}(\theta_1,\theta_2,\theta_3)$ up to two local operations $R^{01}_y(-\frac{\pi}{2})\otimes I_3$ and $R^{01}_y(\frac{\pi}{2})\otimes I_3$. Here
\begin{eqnarray}\label{eq28}
\begin{aligned}
\widetilde{A}(\theta_1,\theta_2,\theta_3)
=&R^{01}_y(-\frac{\pi}{2})\otimes I_3 \cdot A(\theta_1,\theta_2,\theta_3)\\& \cdot R^{01}_y(\frac{\pi}{2})\otimes I_3\\
=&\textrm{exp}
[-\textrm{i}\sigma_z^{01}\otimes(\theta_1 \sigma_z^{01}+\theta_2 \sigma_z^{02}+\theta_3 I_3)].
\end{aligned}
\end{eqnarray}
Observe that $\widetilde{A}(\theta_1,\theta_2,\theta_3)$ is a conditional control gate, which performs $R_z^{01}[\theta(m)]$ operations on the first (target) qutrit when the second (control) qutrit is $|m\rangle$, where $\theta(m)$ depends on $|m\rangle$.
By calculation, we obtain
\begin{eqnarray}\label{eq29}
\begin{aligned}
\widetilde{A}(\theta_1,\theta_2,\theta_3)
=&\textrm{GCX}_2^1(1\rightarrow X^{01}) \cdot R_z^{01}(2\theta_{1}+\theta_{2})\otimes I_3\\&
\cdot \textrm{GCX}_2^1(1\rightarrow X^{01}) \cdot \textrm{GCX}_2^1(2\rightarrow X^{01}) \\&
\cdot R_z^{01}(\theta_{1}+2\theta_{2})\otimes I_3 \cdot \textrm{GCX}_2^1(2\rightarrow X^{01})\\&
\cdot R_z^{01}(2\theta_{3}-\theta_{1}-\theta_{2})\otimes I_3.
\end{aligned}
\end{eqnarray}
Based on Eq. (\ref{eq28}) and Theorem 2, $A(\theta_1,\theta_2,\theta_3)$ can be recovered and further simplified as
\begin{eqnarray}\label{eq30}
\begin{aligned}
A(\theta_1,\theta_2,\theta_3)=&R^{01}_y(\frac{\pi}{2})\otimes I_3 \cdot \widetilde{A}(\theta_1,\theta_2,\theta_3)\\& \cdot R^{01}_y(-\frac{\pi}{2})\otimes I_3\\
=&R^{01}_y(\frac{\pi}{2})\otimes I_3 \cdot \textrm{GCX}_2^1(1\rightarrow X^{01})\\&
 \cdot R_z^{01}(2\theta_{1}+\theta_{2})\otimes I_3\\&
\cdot \textrm{GCX}_2^1(0\rightarrow X^{01}) \cdot X^{01} \otimes I_3 \\&
\cdot R_z^{01}(\theta_{1}+2\theta_{2})\otimes I_3 \\& \cdot \textrm{GCX}_2^1(2\rightarrow X^{01})\\&
\cdot R_z^{01}(2\theta_{3}-\theta_{1}-\theta_{2})\otimes I_3 \\& \cdot R^{01}_y(-\frac{\pi}{2})\otimes I_3.
\end{aligned}
\end{eqnarray}
Here single-qutrit gate $X^{01}$ can be decomposed into three rotation gates as follows:
\begin{eqnarray}\label{eq31}
\begin{aligned}
X^{01}=e^{\textrm{i}\frac{\pi}{3}} R^{01}_x(\pi)\cdot R^{02}_z(-\frac{2\pi}{3})\cdot R^{01}_z(\frac{\pi}{3}),
\end{aligned}
\end{eqnarray}
%

\begin{figure} [htbp]
  \centering
  \includegraphics[width=8.6cm]{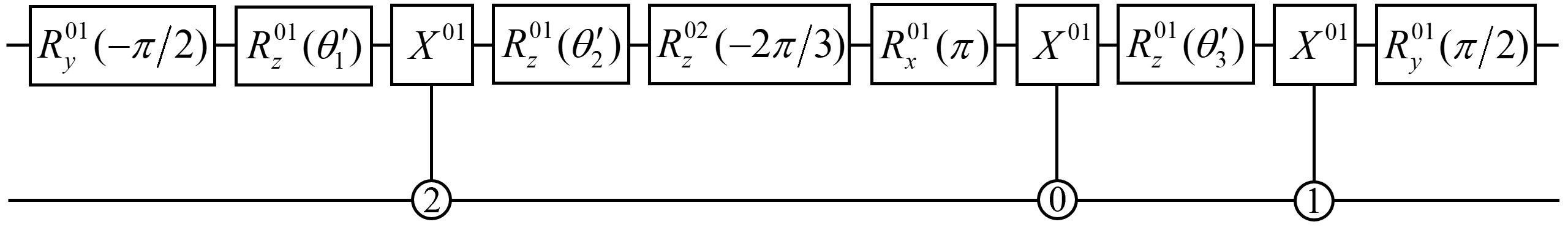}
  \caption{Synthesis of $A(\theta_1,\theta_2,\theta_3)$. Here $\theta_1'=2\theta_{3}-\theta_{1}-\theta_{2}$, $\theta_2'=\theta_{1}+2\theta_{2}+\frac{\pi}{3}$, and $\theta_3'=2\theta_{1}+\theta_{2}$. A global phase $e^{\textrm{i}\frac{\pi}{3}}$ is missing here as it does not affect our argument.}
  \label{Fig.4}
\end{figure}

Substituting Eq. (\ref{eq31}) into Eq. (\ref{eq30}) results in a quantum circuit to implement $A(\theta_1,\theta_2,\theta_3)$, which contains 3 GCX gates and 7 single-qutrit $x$-, $y$-, and $z$-axes type rotations as shown in Fig. \ref{Fig.4}.

\textbf{Synthesis of two-qutrit operation \bm{$A_{1,2}^{(1)}$}}.
Based on Eq. (\ref{eq16}),  $A_{1,2}^{(1)}$ can be given by
\begin{eqnarray}\label{eq32}
\begin{aligned}
A^{(1)}(\theta_4,\theta_5,\theta_6)=&\textrm{exp}[-\textrm{i}(\theta_4 \sigma_x^{12}\otimes \sigma_z^{01}+\theta_5 \sigma_x^{12}\otimes \sigma_z^{02}\\&+\theta_6 \sigma_x^{12}\otimes I_3)].
\end{aligned}
\end{eqnarray}

\begin{figure} [htbp]
  \centering
  \includegraphics[width=8.6cm]{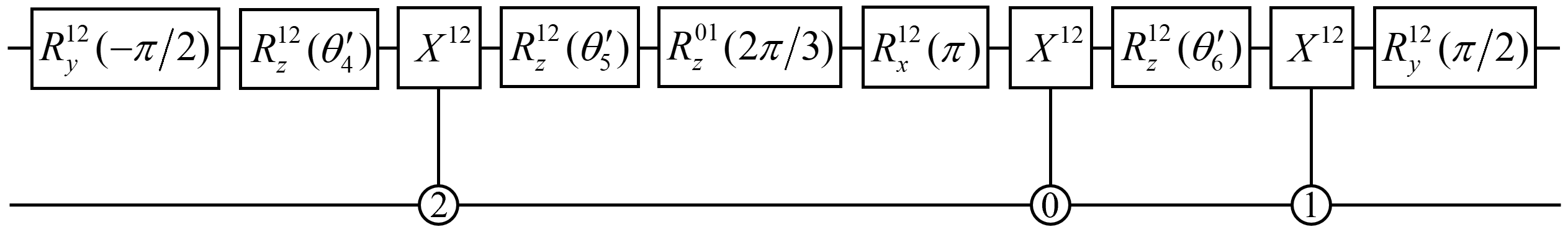}
  \caption{Synthesis of $A^{(1)}(\theta_4,\theta_5,\theta_6)$. Here $\theta_4'=2\theta_{6}-\theta_{4}-\theta_{5}$, $\theta_5'=\theta_{4}+2\theta_{5}+\frac{\pi}{3}$, and $\theta_6'=2\theta_{4}+\theta_{5}$. We ignore the overall phase $e^{\textrm{i}\frac{\pi}{3}}$ here.}
  \label{Fig.5}
\end{figure}

With an argument similar to the synthesis of $A(\theta_1,\theta_2,\theta_3)$, we have the following decomposition
\begin{eqnarray}\label{eq33}
\begin{aligned}
A^{(1)}(\theta_4,\theta_5,\theta_6)=&R^{12}_y(\frac{\pi}{2})\otimes I_3 \cdot \textrm{GCX}_2^1(1\rightarrow X^{12})\\&
 \cdot R_z^{12}(2\theta_{4}+\theta_{5})\otimes I_3\\&
\cdot \textrm{GCX}_2^1(0\rightarrow X^{12}) \cdot X^{12} \otimes I_3 \\&
\cdot R_z^{12}(\theta_{4}+2\theta_{5})\otimes I_3 \\& \cdot \textrm{GCX}_2^1(2\rightarrow X^{12})\\&
\cdot R_z^{12}(2\theta_{6}-\theta_{4}-\theta_{5})\otimes I_3 \\& \cdot R^{12}_y(-\frac{\pi}{2})\otimes I_3.
\end{aligned}
\end{eqnarray}
Here $X^{12}$ can be simulated by
\begin{eqnarray}\label{eq34}
\begin{aligned}
X^{12}=e^{\textrm{i}\frac{\pi}{3}} R^{12}_x(\pi)\cdot R^{01}_z(\frac{2\pi}{3})\cdot R^{12}_z(\frac{\pi}{3}).
\end{aligned}
\end{eqnarray}
Therefore, as shown in Fig. \ref{Fig.5}, 3 GCX gates and 7 single-qutrit $x$-, $y$-, and $z$-axes type rotations are sufficient to simulate $A^{(1)}(\theta_4,\theta_5,\theta_6)$.

\textbf{Synthesis of two-qutrit operation $\mathbf{\emph{A}^{(2)}}$}.
Based on Eq. (\ref{eq20}),  $A^{(2)}$ is diagonal and has the following form
\begin{eqnarray}\label{eq35}
\begin{aligned}
A^{(2)}(\theta_7,\theta_8,\theta_9)=&\textrm{exp}[-\textrm{i}(\theta_7 \sigma_z^{12}\otimes I_3+\theta_8 \sigma_z^{12}\otimes D\\&+\theta_9 \sigma_z^{12}\otimes \sigma_z^{12})].
\end{aligned}
\end{eqnarray}

\begin{figure} [htbp]
  \centering
  \includegraphics[width=6.8cm]{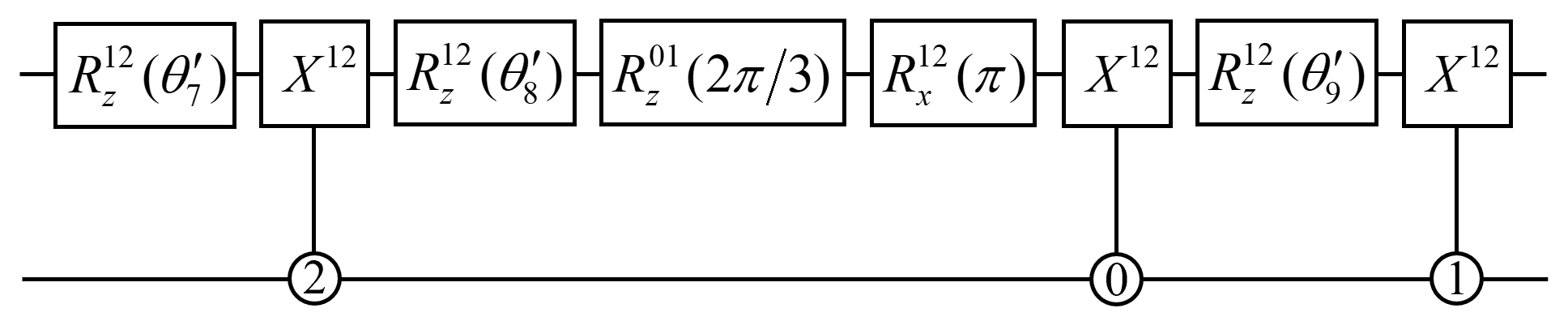}
  \caption{Synthesis of $A^{(2)}(\theta_7,\theta_8,\theta_9)$. Here $\theta_7'=2\theta_{7}-2\theta_{8}$, $\theta_8'=2\theta_{8}+\theta_{9}+\frac{\pi}{3}$, and $\theta_9'=2\theta_{8}-\theta_{9}$. We omit the global phase factor $e^{\textrm{i}\frac{\pi}{3}}$ here.}
  \label{Fig.6}
\end{figure}

By calculation, $A^{(2)}(\theta_7,\theta_8,\theta_9)$ can be decomposed as follow
\begin{eqnarray}\label{eq36}
\begin{aligned}
A^{(2)}(\theta_7,\theta_8,\theta_9)=& \textrm{GCX}_2^1(1\rightarrow X^{12})\\&
 \cdot R_z^{12}(2\theta_{8}-\theta_{9})\otimes I_3\\&
\cdot \textrm{GCX}_2^1(0\rightarrow X^{12}) \cdot X^{12} \otimes I_3 \\&
\cdot R_z^{12}(2\theta_{8}+\theta_{9})\otimes I_3 \\& \cdot \textrm{GCX}_2^1(2\rightarrow X^{12})\\&
\cdot R_z^{12}(2\theta_{7}-2\theta_{8})\otimes I_3.
\end{aligned}
\end{eqnarray}
As a consequence of substituting Eq. (\ref{eq34}) into Eq. (\ref{eq36}), the structure of $A^{(2)}(\theta_7,\theta_8,\theta_9)$ contains 3 GCXs and 5 $x$-, $z$-axes type rotations, see Fig. \ref{Fig.6}.

\textbf{Synthesis of two-qutrit operation \bm{$A^{(3)}_{1,2}$}}.
Based on Eq. (\ref{eq24}), $A^{(3)}_{1,2}$ is of the form
\begin{eqnarray}\label{eq37}
\begin{aligned}
A^{(3)}(\theta_{10},\theta_{11},\theta_{12})=&\textrm{exp}[-\textrm{i} (\theta_{10} D\otimes I_3+\theta_{11} D\otimes D\\&+\theta_{12} D\otimes\sigma_z^{12})].
\end{aligned}
\end{eqnarray}

Firstly $A^{(3)}(\theta_{10},\theta_{11},\theta_{12})$ can be decomposed into the following simple gates
\begin{eqnarray}\label{eq38}
\begin{aligned}
A^{(3)}(\theta_{10},\theta_{11},\theta_{12})=
& \textrm{GCX}_1^2(0\rightarrow X^{02}) \\&
\cdot I_3 \otimes R_z^{02}(-2\theta_{12}-\frac{4\theta_{11}}{3}) \\&
\cdot \textrm{GCX}_1^2(0\rightarrow X^{02}) \\&
\cdot  \textrm{GCX}_1^2(0\rightarrow X^{01})\\&
\cdot I_3 \otimes R_z^{01}(2\theta_{12}-\frac{4\theta_{11}}{3})\\&
\cdot \textrm{GCX}_1^2(0\rightarrow X^{01}) \\&
\cdot \textrm{exp}
[-\textrm{i}(\theta_{10}-\frac{\theta_{11}}{3})D]\otimes I_3.
\end{aligned}
\end{eqnarray}
Returning $\textrm{span}\{I_3,D,\sigma_z^{12}\}$ to $\textrm{span}\{\sigma_z^{01},\sigma_z^{02},I_3\}$ as explained in Eq. (\ref{eq18}), we can obtain that
\begin{eqnarray}\label{eq39}
\begin{aligned}
\textrm{exp}[-\textrm{i}(\theta_{10}-\frac{\theta_{11}}{3})D]=
e^{\textrm{i}\frac{\theta'_{10}}{4}}
R_z^{01}(\theta'_{10})\cdot R_z^{02}(\theta'_{10}),
\end{aligned}
\end{eqnarray}
where $\theta'_{10}=\frac{4\theta_{10}}{3}-\frac{4\theta_{11}}{9}$.
As $R_z^{ij}(\theta)\otimes I_3$ is commutable with both $\textrm{GCX}_1^2(0\rightarrow X^{01})$ and $\textrm{GCX}_1^2(0\rightarrow X^{02})$, substituting Eq. (\ref{eq39}) into (\ref{eq38}) induces
\begin{eqnarray}\label{eq40}
\begin{aligned}
A^{(3)}(\theta_{10},\theta_{11},\theta_{12})=& e^{\textrm{i}\frac{\theta'_{10}}{4}} \textrm{GCX}_1^2(0\rightarrow X^{02})\\&
\cdot R_z^{02}(\theta'_{10}) \otimes R_z^{02}(\theta'_{11})\\&
\cdot \textrm{GCX}_1^2(0\rightarrow X^{02}) \\&
\cdot \textrm{GCX}_1^2(0\rightarrow X^{01}) \\&
\cdot R_z^{01}(\theta'_{10}) \otimes R_z^{01}(\theta'_{12}) \\&
\cdot \textrm{GCX}_1^2(0\rightarrow X^{01}),
\end{aligned}
\end{eqnarray}
where $\theta_{11}'=-2\theta_{12}-\frac{4\theta_{11}}{3}$ and $\theta_{12}'=2\theta_{12}-\frac{4\theta_{11}}{3}$.
Thus, up to a global phase $e^{\textrm{i}\frac{\theta'_{10}}{4}}$, 4 GCX gates and 4 $z$-axes type rotations suffice to build a quantum circuit of $A^{(3)}(\theta_{10},\theta_{11},\theta_{12})$, see Fig. \ref{Fig.7}.
In fact, following Fig. \ref{1b}, two adjacent GCX gates in Eq. (\ref{eq40}) can be combined into a CINC gate.

\begin{figure} [htbp]
  \centering
  \includegraphics[width=8.6cm]{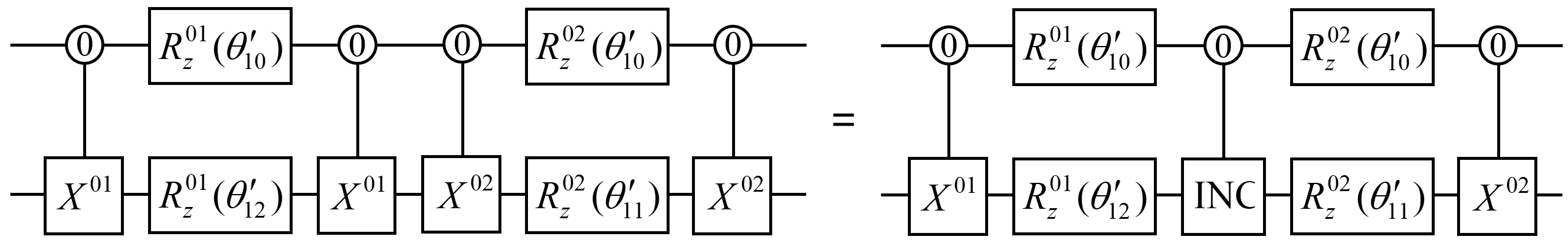}
  \caption{Synthesis of $A^{(3)}(\theta_{10},\theta_{11},\theta_{12})$. Here $\theta'_{10}=\frac{4\theta_{10}}{3}-\frac{4\theta_{11}}{9}$, $\theta_{11}'=-2\theta_{12}-\frac{4\theta_{11}}{3}$, and $\theta_{12}'=2\theta_{12}-\frac{4\theta_{11}}{3}$.}
  \label{Fig.7}
\end{figure}

\textbf{Synthesis of two-qutrit operation \bm{$\overline{A}^{(3)}_{1,2}$}}.
Based on Eq. (\ref{eq25}), $\overline{A}^{(3)}_{1,2}$ can be written in the following form and has the following decomposition
\begin{eqnarray}\label{eq41}
\begin{aligned}
\overline{A}^{(3)}(\theta_{13},\theta_{14},\theta_{15})
=&\textrm{exp}[-\textrm{i} (\theta_{13} \overline{D}\otimes I_3+\theta_{14} \overline{D}\otimes \overline{D}\\&+\theta_{15} \overline{D}\otimes\sigma_z^{01})]\\
=&\textrm{GCX}_1^2(2\rightarrow X^{02}) \\&
\cdot I_3 \otimes R_z^{02}(\frac{8\theta_{14}}{3}) \\&
\cdot \textrm{GCX}_1^2(2\rightarrow X^{02}) \\&
\cdot \textrm{GCX}_1^2(2\rightarrow X^{01})\\&
\cdot I_3 \otimes R_z^{01}(-2\theta_{15}-\frac{4\theta_{14}}{3})\\&
\cdot \textrm{GCX}_1^2(2\rightarrow X^{01}) \\&
\cdot \textrm{exp}
[-\textrm{i}(\theta_{13}-\frac{\theta_{14}}{3})\overline{D}]\otimes I_3.
\end{aligned}
\end{eqnarray}

\begin{figure} [htbp]
  \centering
  \includegraphics[width=8.6cm]{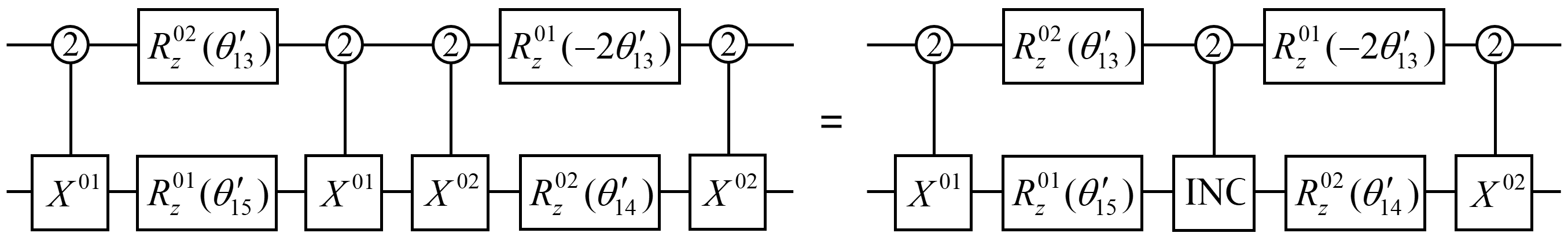}
  \caption{Synthesis of $\overline{A}^{(3)}(\theta_{13},\theta_{14},\theta_{15})$. Here $\theta'_{13}=\frac{4\theta_{13}}{3}-\frac{4\theta_{14}}{9}$, $\theta_{14}'=\frac{8\theta_{14}}{3}$, and $\theta_{15}'=-2\theta_{15}-\frac{4\theta_{14}}{3}$. The quantum circuit on the right is obtained by applying Fig. \ref{1b}.}
  \label{Fig.8}
\end{figure}

Expressing $\overline{D}$ in terms of $\sigma_z^{01}$, $\sigma_z^{02}$, and $I_3$, we get
\begin{eqnarray}\label{eq42}
\begin{aligned}
\textrm{exp}[-\textrm{i}(\theta_{13}-
\frac{\theta_{14}}{3})\overline{D}]=
e^{\textrm{i}\frac{\theta'_{13}}{4}}
R_z^{01}(\theta'_{13})\cdot R_z^{02}(-2\theta'_{13}),
\end{aligned}
\end{eqnarray}
where $\theta'_{13}=\frac{4\theta_{13}}{3}-\frac{4\theta_{14}}{9}$.
Since $R_z^{ij}(\theta)\otimes I_3$ is commutable with both $\textrm{GCX}_1^2(2\rightarrow X^{01})$ and $\textrm{GCX}_1^2(2\rightarrow X^{02})$, substituting Eq. (\ref{eq42}) into (\ref{eq41}) results in
\begin{eqnarray}\label{eq43}
\begin{aligned}
\overline{A}^{(3)}(\theta_{13},\theta_{14},\theta_{15})=& e^{\textrm{i}\frac{\theta'_{13}}{4}}
\textrm{GCX}_1^2(2\rightarrow X^{02}) \\&
\cdot R_z^{02}(-2\theta'_{13}) \otimes R_z^{02}(\theta'_{14}) \\&
\cdot \textrm{GCX}_1^2(2\rightarrow X^{02}) \\&
\cdot\textrm{GCX}_1^2(2\rightarrow X^{01})\\&
\cdot R_z^{01}(\theta'_{13}) \otimes R_z^{01}(\theta'_{15})\\&
\cdot \textrm{GCX}_1^2(2\rightarrow X^{01}),
\end{aligned}
\end{eqnarray}
where $\theta_{14}'=\frac{8\theta_{14}}{3}$ and $\theta_{15}'=-2\theta_{15}-\frac{4\theta_{14}}{3}$.
Thus, similar to Fig. \ref{Fig.7}, 4 GCX gates and 4 $z$-axes type rotations are sufficient to synthesize $\overline{A}^{(3)}(\theta_{13},\theta_{14},\theta_{15})$ up to a global phase $e^{\textrm{i}\frac{\theta'_{13}}{4}}$, see Fig. \ref{Fig.8}.

\textbf{Synthesis of the arbitrary two-qutrit gate}. 
Embedding Fig. \ref{Fig.4}, Fig. \ref{Fig.5}, Fig. \ref{Fig.6}, Fig. \ref{Fig.7}, and Fig. \ref{Fig.8} in Fig. \ref{Fig.3}, one can find that the synthesis of a generic two-qutrit gate requires 21 universal two-qutrit gates (or 25 GCX gates) at most.
In this process, the rightmost GCXs of Fig. \ref{Fig.4} and Fig. \ref{Fig.5} have been absorbed into the block diagonal matrices occurring in the proof of Corollary 1, see Appendix \ref{Appendix.A} for details.
Hence, our Cartan-decomposition-based program can be used to simulate arbitrary two-qutrit gates in terms of GCXs and single-qutrit $R_\varphi^{ij}(\theta)$ gates.

\section{Recursive decomposition of arbitrary multi-qutrit gates}\label{sec4}

The program described in Sec. \ref{sec3} can be extended to synthesize an arbitrary $n$-qutrit gate.

\subsection{Cartan decomposition of Lie group $\mathbf{\emph{U}(3^\emph{n})}$}\label{sec4.1}

Without any loss of generality, let the Lie algebra $\mathfrak{u}(3^{n})$ is spanned as
\begin{eqnarray}\label{eq44}
\begin{aligned}
\mathfrak{u}(3^{n})=\textrm{span}
\{&\sigma_x^{01}\otimes A,\sigma_x^{02}\otimes A,\sigma_x^{12}\otimes A,\sigma_y^{01}\otimes A,\\&\sigma_y^{02}\otimes A,\sigma_y^{12}\otimes A,\sigma_z^{01}\otimes A,\sigma_z^{02}\otimes A,\\&I_3\otimes A|A\in \mathfrak{u}(3^{n-1})\}.
\end{aligned}
\end{eqnarray}

\textbf{Decomposition of \bm{$\mathfrak{u}(3^n)$}}. By calculation, we find that the decomposition of $\mathfrak{u}(3^{n})$ has the following Theorem.

\textbf{Theorem 3.} $\mathfrak{u}(3^n)=\mathfrak{l}(3^n)\oplus \mathfrak{p}(3^n)$ is the Cartan decomposition, where
\begin{eqnarray}\label{eq45}
\begin{aligned}
\mathfrak{l}(3^n)=\textrm{span}\{&\sigma_x^{12}\otimes A,\sigma_y^{12}\otimes A,\sigma_z^{01}\otimes A,\\&\sigma_z^{02}\otimes A,I_3\otimes A|A\in \mathfrak{u}(3^{n-1})\},\\
\mathfrak{p}(3^n)=\textrm{span}\{&\sigma_x^{01}\otimes A,\sigma_x^{02}\otimes A,\sigma_y^{01}\otimes A,\\&\sigma_y^{02}\otimes A|A\in \mathfrak{u}(3^{n-1})\}.
\end{aligned}
\end{eqnarray}
Further, let
\begin{eqnarray}\label{eq46}
\begin{aligned}
&\alpha(3)=\{\textrm{i}I_{3},\textrm{i}\sigma_{z}^{01},\textrm{i}\sigma_{z}^{02}\},\\
&\alpha(3^n)=
\{I_{3}\otimes A,\sigma_{z}^{01}\otimes A,\sigma_{z}^{02}\otimes A|A\in \alpha(3^{n-1})\}.
\end{aligned}
\end{eqnarray}
The Cartan subalgebra of $(\mathfrak{u}(3^n),\mathfrak{l}(3^n))$ can be picked as
\begin{eqnarray}\label{eq47}
\begin{aligned}
\mathfrak{a}(3^n)=\textrm{span}\{\sigma_x^{01}\otimes A|A\in \alpha(3^{n-1})\}\subset\mathfrak{p}(3^n).
\end{aligned}
\end{eqnarray}

\textbf{Proof 3.}\textbf{ Firstly}, we prove that $[\mathfrak{l}(3^n),\mathfrak{l}(3^n)]\subseteq\mathfrak{l}(3^n)$.

Given that any $M_1, M_2\in \mathfrak{l}(3^n)$, then $M_1$ and $M_2$ can be denoted as
\begin{eqnarray}\label{eq48}
\begin{aligned}
M_1=&I_3\otimes A_1+\sigma_z^{01}\otimes A_2+\sigma_z^{02}\otimes A_3+\sigma_x^{12}\otimes A_4\\&+\sigma_y^{12}\otimes A_5
\\=&\left(
  \begin{array}{ccc}
    A_1+A_2+A_3 & \textbf{0} & \textbf{0}\\
    \textbf{0} & A_1-A_2 & A_4-\textrm{i}A_5\\
    \textbf{0} & A_4+\textrm{i}A_5 & A_1-A_3\\
  \end{array}
\right),
\end{aligned}
\end{eqnarray}
\begin{eqnarray}\label{eq49}
\begin{aligned}
M_2=&I_3\otimes B_1+\sigma_z^{01}\otimes B_2+\sigma_z^{02}\otimes B_3+\sigma_x^{12}\otimes B_4\\&+\sigma_y^{12}\otimes B_5
\\=&\left(
  \begin{array}{ccc}
    B_1+B_2+B_3 & \textbf{0} & \textbf{0}\\
    \textbf{0} & B_1-B_2 & B_4-\textrm{i}B_5\\
    \textbf{0} & B_4+\textrm{i}B_5 & B_1-B_3\\
  \end{array}
\right).
\end{aligned}
\end{eqnarray}
Here $A_i, B_j\in\mathfrak{u}(3^{n-1})$, i.e., $A_i$ and $B_j$ are $3^{n-1}\times3^{n-1}$ skew-Hermitian matrices. $M_1$ and $M_2$ obey the commutator relations
\begin{eqnarray}\label{eq50}
\begin{aligned}
[M_1,M_2]&= M_1\cdot M_2 - M_2\cdot M_1\\
&=\left(
  \begin{array}{ccc}
    M_{11} & \textbf{0} & \textbf{0}\\
    \textbf{0} & M_{22} & M_{23}\\
    \textbf{0} & M_{32} & M_{33}\\
  \end{array}
\right),
\end{aligned}
\end{eqnarray}
where $M_{11}^{\dagger}=-M_{11}$, $M_{22}^{\dagger}=-M_{22}$, $M_{33}^{\dagger}=-M_{33}$, and $M_{23}^{\dagger}=-M_{32}$.
Here $\dagger$ denotes a conjugate transposition.

By setting
\begin{eqnarray}\label{eq51}
\begin{aligned}
&X_1=\frac{M_{11}+M_{22}+M_{33}}{3},\\&
X_2=\frac{M_{11}-2M_{22}+M_{33}}{3},\\&
X_3=\frac{M_{11}+M_{22}-2M_{33}}{3},\\&
X_4=\frac{M_{23}+M_{32}}{3},\\&
X_5=\textrm{i}\frac{M_{23}-M_{32}}{3},
\end{aligned}
\end{eqnarray}
$[M_1,M_2]$ shown in Eq. (\ref{eq50}) can be rewritten as
\begin{eqnarray}\label{eq52}
\begin{aligned}
\left(
  \begin{array}{ccc}
    X_1+X_2+X_3 & \textbf{0} & \textbf{0}\\
    \textbf{0} & X_1-X_2 & X_4-\textrm{i}X_5\\
    \textbf{0} & X_4+\textrm{i}X_5 & X_1-X_3\\
  \end{array}
\right).
\end{aligned}
\end{eqnarray}
This form is the same as that presented in Eqs. (\ref{eq48}) and (\ref{eq49}) due to $X_{i}^{\dagger}=-X_{i}$.
Hence, $[M_1,M_2]\in\mathfrak{l}(3^n)$, and than $[\mathfrak{l}(3^n),\mathfrak{l}(3^n)]\subseteq\mathfrak{l}(3^n)$.

\textbf{Secondly}, we prove that $[\mathfrak{p}(3^n),\mathfrak{p}(3^n)]\subseteq\mathfrak{l}(3^n)$.

Suppose $M_3,M_4\in\mathfrak{p}(3^n)$, then $M_3$ and $M_4$ have the following generic forms
\begin{eqnarray}\label{eq53}
\begin{aligned}
M_3=&\sigma_x^{01}\otimes C_1+\sigma_x^{02}\otimes C_2+\sigma_y^{01}\otimes C_3+\sigma_y^{02}\otimes C_4
\\=&\left(
  \begin{array}{ccc}
    \textbf{0} & C_1-\textrm{i}C_3 & C_2-\textrm{i}C_4\\
    C_1+\textrm{i}C_3 & \textbf{0} & \textbf{0}\\
    C_2+\textrm{i}C_4 & \textbf{0} & \textbf{0}\\
  \end{array}
\right),
\end{aligned}
\end{eqnarray}
\begin{eqnarray}\label{eq54}
\begin{aligned}
M_4=&\sigma_x^{01}\otimes D_1+\sigma_x^{02}\otimes D_2+\sigma_y^{01}\otimes D_3+\sigma_y^{02}\otimes D_4
\\=&\left(
  \begin{array}{ccc}
    \textbf{0} & D_1-\textrm{i}D_3 & D_2-\textrm{i}D_4\\
    D_1+\textrm{i}D_3 & \textbf{0} & \textbf{0}\\
    D_2+\textrm{i}D_4 & \textbf{0} & \textbf{0}\\
  \end{array}
\right),
\end{aligned}
\end{eqnarray}
where $C_k,D_l\in\mathfrak{u}(3^{n-1})$.
By multiplication of the block matrices, we get
\begin{eqnarray}\label{eq55}
\begin{aligned}
[M_3,M_4]
=\left(
  \begin{array}{ccc}
    \overline{M}_{11} & \textbf{0} & \textbf{0}\\
    \textbf{0} & \overline{M}_{22} & \overline{M}_{23}\\
    \textbf{0} & \overline{M}_{32} & \overline{M}_{33}\\
  \end{array}
\right),
\end{aligned}
\end{eqnarray}
where $\overline{M}_{11}^{\dagger}=-\overline{M}_{11}$, $\overline{M}_{22}^{\dagger}=-\overline{M}_{22}$, $\overline{M}_{33}^{\dagger}=-\overline{M}_{33}$, and $\overline{M}_{23}^{\dagger}=-\overline{M}_{32}$.
Therefore, in conjunction with the previous argument, it is clear that $[\mathfrak{p}(3^n),\mathfrak{p}(3^n)]\subseteq\mathfrak{l}(3^n)$.

\textbf{Thirdly}, we prove that $[\mathfrak{l}(3^n),\mathfrak{p}(3^n)]\subseteq\mathfrak{p}(3^n)$.

Calculate $[M_1,M_3]$ to get
\begin{eqnarray}\label{eq56}
\begin{aligned}
&[M_1,M_3]
=\left(
  \begin{array}{ccc}
    \textbf{0} & M_{12} & M_{13}\\
    M_{21} & \textbf{0} & \textbf{0}\\
    M_{31} & \textbf{0} & \textbf{0}\\
  \end{array}
\right),
\end{aligned}
\end{eqnarray}
where $M_{12}^{\dagger}=-M_{21}$, $M_{13}^{\dagger}=-M_{31}$.
Setting
\begin{eqnarray}\label{eq57}
\begin{aligned}
&Y_1=\frac{M_{12}+M_{21}}{2},\\&
Y_2=\frac{M_{13}+M_{31}}{2},\\&
Y_3=\textrm{i}\frac{M_{12}-M_{21}}{2},\\&
Y_4=\textrm{i}\frac{M_{13}-M_{31}}{2},
\end{aligned}
\end{eqnarray}
Eq. (\ref{eq56}) is written as follows
\begin{eqnarray}\label{eq58}
\begin{aligned}
&[M_1,M_3]=\left(
  \begin{array}{ccc}
    \textbf{0} & Y_1-\textrm{i}Y_3 & Y_2-\textrm{i}Y_4\\
    Y_1+\textrm{i}Y_3 & \textbf{0} & \textbf{0}\\
    Y_2+\textrm{i}Y_4 & \textbf{0} & \textbf{0}\\
  \end{array}
\right).
\end{aligned}
\end{eqnarray}
Since $Y_{i}^{\dagger}=-Y_{i}$, Eq. (\ref{eq58}) has the same form as Eq. (\ref{eq53}).
Therefore, $[M_1,M_3]\in\mathfrak{p}(3^n)$, and $[\mathfrak{l}(3^n),\mathfrak{p}(3^n)]\subseteq\mathfrak{p}(3^n)$.

\textbf{Lastly}, we prove that $\mathfrak{a}(3^n)$ described by Eq. (\ref{eq47}) is the Cartan subalgebra of $(\mathfrak{u}(3^n),\mathfrak{l}(3^n))$.
Obviously, any element in $\mathfrak{a}(3^n)$ commutes to each other. In the following, we prove that $\mathfrak{a}(3^n)$ is maximally Abelian contained in $\mathfrak{p}(3^n)$.

Given that any $A\in\mathfrak{p}(3^n)$, $A$ can be expressed as
\begin{eqnarray}\label{eq59}
A=\sigma_x^{01}\otimes A_1+\sigma_y^{01}\otimes A_2+\sigma_x^{02}\otimes A_3+\sigma_y^{02}\otimes A_4,
\end{eqnarray}
where $A_i\in\mathfrak{u}(3^{n-1})$.
Assume that $A$ and $\mathfrak{a}(3^n)$ are commutable, i.e., $[A, \mathfrak{a}(3^n)]=\textbf{0}$, let us prove that $A\in\mathfrak{a}(3^n)$.
Since $\sigma_x^{01}\otimes I_{3^{n-1}}\in\mathfrak{a}(3^n)$, we have
\begin{eqnarray}\label{eq60}
A\cdot\sigma_x^{01}\otimes I_{3^{n-1}}=\sigma_x^{01}\otimes I_{3^{n-1}}\cdot A.
\end{eqnarray}
This leads to $A_2=A_3=A_4=\mathbf{0}$, and then $A=\sigma_x^{01}\otimes A_1$.

Thus, $[A, \mathfrak{a}(3^n)]=\textbf{0}$ implies that $[A_1, \alpha(3^{n-1})]=\textbf{0}$.
By Lemma (see Appendix \ref{Appendix.B}), it follows that $A_1\in\textrm{span}\{\alpha(3^{n-1})\}$, and then $A\in\mathfrak{a}(3^n)$.
Therefore, $\mathfrak{a}(3^n)$ is maximally Abelian in $\mathfrak{p}(3^n)$. $\hfill\blacksquare$

\textbf{Decomposition of \bm{$\mathfrak{l}(3^n)$}}.
Note that $\mathfrak{l}(3^n)$ is isomorphic to $\mathfrak{u}(3^{n-1})\oplus \mathfrak{u}(2\times3^{n-1})$.
Similarly to the two-qutrit case, $\mathfrak{l}(3^{n})$ can be further decomposition.

\textbf{Theorem 4.} $\mathfrak{l}(3^n)=\mathfrak{l}_1(3^n)\oplus \mathfrak{p}_1(3^n)$ is a Cartan decomposition of Lie subalgebra $\mathfrak{l}(3^n)$ with
\begin{eqnarray}\label{eq61}
\begin{aligned}
&\mathfrak{l}_1(3^{n})=\textrm{span}\{\sigma_z^{01}\otimes A,\sigma_z^{02}\otimes A,I_3\otimes A|A\in \mathfrak{u}(3^{n-1})\},\\
&\mathfrak{p}_1(3^{n})=\textrm{span}\{\sigma_x^{12}\otimes A,\sigma_y^{12}\otimes A,|A\in \mathfrak{u}(3^{n-1})\},\\
&\mathfrak{a}_1(3^n)=\textrm{span}\{\sigma_x^{12}\otimes A|A\in \alpha(3^{n-1})\}.
\end{aligned}
\end{eqnarray}
Here $\mathfrak{a}_1(3^n)$ is the Cartan subalgebra of $(\mathfrak{l}(3^n),\mathfrak{l}_1(3^n))$.

\textbf{Proof 4.} \textbf{Firstly}, we prove that $[\mathfrak{l}_1(3^n),\mathfrak{l}_1(3^n)]\subseteq\mathfrak{l}_1(3^n)$.

For any $A_1, A_2\in\mathfrak{u}(3^{n-1})$, we have the following relations
\begin{eqnarray}\label{eq62}
\begin{aligned}
&[I_3\otimes A_1,\sigma_z^{0i}\otimes A_2]=\sigma_z^{0i}\otimes [A_1,A_2],
\\&
[\sigma_z^{0i}\otimes A_1,\sigma_z^{0j}\otimes A_2]=\frac{I_3+\sigma_z^{0i}+\sigma_z^{0j}}{3}\otimes [A_1,A_2],
\\&
[\sigma_z^{0i}\otimes A_1,\sigma_z^{0i}\otimes A_2]=\frac{2I_3+2\sigma_z^{0j}-\sigma_z^{0i}}{3}\otimes [A_1,A_2],
\end{aligned}
\end{eqnarray}
where $i,j\in\{0,1\}$ and $i\neq j$. Eq. (\ref{eq62}) and  $[A_1,A_2]\in\mathfrak{u}(3^{n-1})$ indicate that $[\mathfrak{l}_1(3^n),\mathfrak{l}_1(3^n)]\subseteq\mathfrak{l}_1(3^n)$.

\textbf{Secondly}, we prove that $[\mathfrak{p}_1(3^n),\mathfrak{p}_1(3^n)]\subseteq\mathfrak{l}_1(3^n)$.

A calculation yields the following relations
\begin{eqnarray}\label{eq63}
\begin{aligned}
&[\sigma_x^{12}\otimes A_1,\sigma_y^{12}\otimes A_2]=\sigma_z^{12}\otimes \textrm{i}(A_1A_2+A_2A_1),\\&
[\sigma_x^{12}\otimes A_1,\sigma_x^{12}\otimes A_2]=\frac{2I_3-\sigma_z^{01}-\sigma_z^{02}}{3}\otimes [A_1,A_2],\\&
[\sigma_y^{12}\otimes A_1,\sigma_y^{12}\otimes A_2]=\frac{2I_3-\sigma_z^{01}-\sigma_z^{02}}{3}\otimes [A_1,A_2].
\end{aligned}
\end{eqnarray}
Eq. (\ref{eq63}) and $\sigma_{z}^{12}=\sigma_{z}^{02}-\sigma_{z}^{01}$ results in $[\mathfrak{p}_1(3^n),\mathfrak{p}_1(3^n)]\subseteq\mathfrak{l}_1(3^n)$.

\textbf{Thirdly}, we prove that $[\mathfrak{l}_1(3^n),\mathfrak{p}_1(3^n)]\subseteq\mathfrak{p}_1(3^n)$.

Any $M_1\in \mathfrak{l}_1(3^n)$ and $M_2\in \mathfrak{p}_1(3^n)$ can write as
\begin{eqnarray}\label{eq64}
\begin{aligned}
M_1=&I_3\otimes B_1+\sigma_z^{01}\otimes B_2+\sigma_z^{02}\otimes B_3
\\=&\left(
  \begin{array}{ccc}
    B_1+B_2+B_3 & \textbf{0} & \textbf{0}\\
    \textbf{0} & B_1-B_2 & \textbf{0}\\
    \textbf{0} & \textbf{0} & B_1-B_3\\
  \end{array}
\right),
\end{aligned}
\end{eqnarray}
\begin{eqnarray}\label{eq65}
\begin{aligned}
M_2=&\sigma_x^{12}\otimes C_1+\sigma_y^{12}\otimes C_2
\\=&\left(
  \begin{array}{ccc}
    \textbf{0} & \textbf{0} & \textbf{0}\\
    \textbf{0} & \textbf{0} & C_1-\textrm{i}C_2\\
    \textbf{0} & C_1+\textrm{i}C_2 & \textbf{0}\\
  \end{array}
\right).
\end{aligned}
\end{eqnarray}
Here $B_k, C_l\in\mathfrak{u}(3^{n-1})$. $M_1$ and $M_2$ satisfy the following relation
\begin{eqnarray}\label{eq66}
\begin{aligned}
&[M_1,M_2]=\left(
  \begin{array}{ccc}
    \textbf{0} & \textbf{0} & \textbf{0}\\
    \textbf{0} & \textbf{0} & M_{23}\\
    \textbf{0} & M_{32} & \textbf{0}\\
  \end{array}
\right)
\end{aligned}
\end{eqnarray}
where $M_{23}^{\dagger}=-M_{32}$. By setting
\begin{eqnarray}\label{eq67}
\begin{aligned}
&X_1=\frac{M_{23}+M_{32}}{2}\in \mathfrak{u}(3^{n-1}), \\&
X_2=\textrm{i}\frac{M_{23}-M_{32}}{2}\in \mathfrak{u}(3^{n-1}),
\end{aligned}
\end{eqnarray}
Eq. (\ref{eq66}) can be written as
\begin{eqnarray}\label{eq68}
[M_1,M_2]=\sigma_x^{12}\otimes X_1+\sigma_y^{12}\otimes X_2.
\end{eqnarray}
This means that $[M_1,M_2]\in\mathfrak{p}_1(3^n)$, and than $[\mathfrak{l}_1(3^n),\mathfrak{p}_1(3^n)]\subseteq\mathfrak{p}_1(3^n)$.

\textbf{Finally}, we prove that $\mathfrak{a}_1(3^n)$ described by Eq. (\ref{eq61}) is a Cartan subalgebra  of the pair $(\mathfrak{l}(3^n),\mathfrak{l}_1(3^n))$.
It is clear that $\mathfrak{a}_1(3^n)$ is Abelian. We now prove that $\mathfrak{a}_1(3^n)$ is maximally Abelian in $\mathfrak{p}_1(3^n)$.

For any $A\in\mathfrak{p}_1(3^n)$, let
\begin{eqnarray}\label{eq69}
A=\sigma_x^{12}\otimes A_1+\sigma_y^{12}\otimes A_2
\end{eqnarray}
where $A_i\in\mathfrak{u}(3^{n-1})$.
Supposing that $A$ and $\mathfrak{a}_1(3^n)$ are commutable, as $\sigma_x^{12}\otimes I_{3^{n-1}}\in\mathfrak{a}(3^n)$, we have
\begin{eqnarray}\label{eq70}
A\cdot\sigma_x^{12}\otimes I_{3^{n-1}}=\sigma_x^{12}\otimes I_{3^{n-1}}\cdot A.
\end{eqnarray}
Thus, $A=\sigma_x^{12}\otimes A_1$.
By assumption, it follows that $A_1$ and $\textrm{span}\{\alpha(3^{n-1})\}$ are commutable.
From Lemma, we have $A\in\mathfrak{a}_1(3^n)$.
Therefore, the theorem follows. $\hfill\blacksquare$

\textbf{Decomposition of \bm{$\mathfrak{l}_1(3^n)$}}.
One can find that $\mathfrak{l}_1(3^{n})$ is isomorphic to $\mathfrak{u}(3^{n-1})\oplus \mathfrak{u}(3^{n-1}) \oplus \mathfrak{u}(3^{n-1})$.
Utilizing Eq. (\ref{eq17}), $\mathfrak{l}_1(3^n)$ is rewritten as
\begin{eqnarray}\label{eq71}
\begin{aligned}
\mathfrak{l}_1(3^n)&=\textrm{span}\{I_3\otimes A,D\otimes A,\sigma_z^{12}\otimes A|A\in \mathfrak{u}(3^{n-1})\}.\\&
=\textrm{span}\{I_3\otimes A,\overline{D}\otimes A,\sigma_z^{01}\otimes A|A\in \mathfrak{u}(3^{n-1})\}.
\end{aligned}
\end{eqnarray}
Then $\mathfrak{l}_1(3^n)$ becomes decomposable and the decomposition complies with the following Theorem.

\textbf{Theorem 5.} $\mathfrak{l}_1(3^n)$ has the following Cartan decomposition
\begin{eqnarray}\label{eq72}
\begin{aligned}
&\mathfrak{l}_1(3^n)=\mathfrak{l}_2(3^n)\oplus \mathfrak{p}_2(3^n),\\
&\mathfrak{l}_2(3^n)=\textrm{span}\{I_3\otimes A,D\otimes A|A\in \mathfrak{u}(3^{n-1})\},\\
&\mathfrak{p}_2(3^n)=\textrm{span}\{\sigma_z^{12}\otimes A|A\in \mathfrak{u}(3^{n-1})\},\\
&\mathfrak{a}_2(3^n)=\textrm{span}\{\sigma_z^{12}\otimes A|A\in \widetilde{\alpha}(3^{n-1})\}\subset\mathfrak{p}_2(3^n).
\end{aligned}
\end{eqnarray}
Here $\mathfrak{a}_2(3^n)$ is the Cartan subalgebra of $(\mathfrak{l}_1(3^n),\mathfrak{l}_2(3^n))$ and
\begin{eqnarray}\label{eq73}
\begin{aligned}
&\widetilde{\alpha}(3)=\{\textrm{i}I_{3},\textrm{i}D,\textrm{i}\sigma_{z}^{12}\},\\
&\widetilde{\alpha}(3^n)=\{I_{3}\otimes A,D\otimes A,\sigma_{z}^{12}\otimes A|A\in \widetilde{\alpha}(3^{n-1})\}.
\end{aligned}
\end{eqnarray}
In addition, there is another Cartan decomposition of $\mathfrak{l}_1(3^n)$ as follows
\begin{eqnarray}\label{eq74}
\begin{aligned}
&\mathfrak{l}_1(3^n)=\overline{\mathfrak{l}_2}(3^n)\oplus \overline{\mathfrak{p}_2}(3^n),\\
&\overline{\mathfrak{l}_2}(3^n)=\textrm{span}\{I_3\otimes A,\overline{D}\otimes A|A\in \mathfrak{u}(3^{n-1})\},\\
&\overline{\mathfrak{p}_2}(3^n)=\textrm{span}\{\sigma_z^{01}\otimes A|A\in \mathfrak{u}(3^{n-1})\}.
\end{aligned}
\end{eqnarray}

\textbf{Proof 5.} Given any $A_1,A_2\in \mathfrak{u}(3^{n-1})$, there are the following commutation relations
\begin{eqnarray}\label{eq75}
\begin{aligned}
&[I_3\otimes A_1,D\otimes A_2]=D\otimes [A_1,A_2],
\end{aligned}
\end{eqnarray}
\begin{eqnarray}\label{eq76}
\begin{aligned}
&[D\otimes A_1,D\otimes A_2]=I_3\otimes [A_1,A_2],
\end{aligned}
\end{eqnarray}
\begin{eqnarray}\label{eq77}
\begin{aligned}
&[I_3\otimes A_1,\sigma_z^{12}\otimes A_2]=\sigma_z^{12}\otimes [A_1,A_2],
\end{aligned}
\end{eqnarray}
\begin{eqnarray}\label{eq78}
\begin{aligned}
&[D\otimes A_1,\sigma_z^{12}\otimes A_2]=-\sigma_z^{12}\otimes [A_1,A_2],
\end{aligned}
\end{eqnarray}
\begin{eqnarray}\label{eq79}
\begin{aligned}
&[\sigma_z^{12}\otimes A_1,\sigma_z^{12}\otimes A_2]=\frac{I_3-D}{2}\otimes [A_1,A_2].
\end{aligned}
\end{eqnarray}
Eqs. (\ref{eq75}) and (\ref{eq76}) indicate that $[\mathfrak{l}_2(3^n),\mathfrak{l}_2(3^n)]\subseteq\mathfrak{l}_2(3^n)$.
Eqs. (\ref{eq77}) and (\ref{eq78}) indicate that $[\mathfrak{l}_2(3^n),\mathfrak{p}_2(3^n)]\subseteq\mathfrak{p}_2(3^n)$.
Eq. (\ref{eq79}) indicates that $[\mathfrak{p}_2(3^n),\mathfrak{p}_2(3^n)]\subseteq\mathfrak{l}_2(3^n)$.
Similarly, Eq. (\ref{eq74}) is found to hold by simply replacing $D$ with $\overline{D}$ and $\sigma_z^{12}$ with $\sigma_z^{01}$ in Eqs. (\ref{eq75}-\ref{eq79}).

Now prove that $\mathfrak{a}_2(3^n)$ is a Cartan subalgebra of the pair $(\mathfrak{l}_1(3^n),\mathfrak{l}_2(3^n))$.
From Eq. (\ref{eq17}), it can be found that
\begin{eqnarray}\label{eq78}
\textrm{span}\{\widetilde{\alpha}(3^n)\}=\textrm{span}\{\alpha(3^n)\}.
\end{eqnarray}
Under Lemma, it follows that $\textrm{span}\{\widetilde{\alpha}(3^{n-1})\}$ is the maximally Abelian in $\mathfrak{u}(3^{n-1})$. Recalling the definition of $\mathfrak{a}_2(3^n)$ results in $\mathfrak{a}_2(3^{n})$ being maximally Abelian in $\mathfrak{p}_2(3^n)$. $\hfill\blacksquare$

\textbf{Decomposition of \bm{$\mathfrak{l}_2(3^n)$} and \bm{$\overline{\mathfrak{l}_2}(3^n)$}}. Note that both $\mathfrak{l}_2(3^n)$ and $\overline{\mathfrak{l}_2}(3^n)$ are isomorphic to $\mathfrak{u}(3^{n-1})\oplus \mathfrak{u}(3^{n-1})$.
It is possible to further distinguish between local contents and the nonlocal among them.

\textbf{Theorem 6.} $\mathfrak{l}_2(3^n)=\mathfrak{l}_3(3^n)\oplus \mathfrak{p}_3(3^n)$ is a Cartan decomposition of $\mathfrak{l}_2(3^n)$ with a Cartan subalgebra $\mathfrak{a}_3(3^n)$, where
\begin{eqnarray}\label{eq81}
\begin{aligned}
&\mathfrak{l}_3(3^n)=\textrm{span}\{I_3\otimes A|A\in \mathfrak{u}(3^{n-1})\},\\
&\mathfrak{p}_3(3^n)=\textrm{span}\{D  \otimes A|A\in \mathfrak{u}(3^{n-1})\},\\
&\mathfrak{a}_3(3^n)=\textrm{span}\{D\otimes A|A\in\widetilde{\alpha}(3^{n-1})\}.
\end{aligned}
\end{eqnarray}
$\overline{\mathfrak{l}_2}(3^n)$ has the following Cartan decomposition
\begin{eqnarray}\label{eq82}
\begin{aligned}
&\overline{\mathfrak{l}_2}(3^n)=\mathfrak{l}_3(3^n)\oplus \overline{\mathfrak{p}_3}(3^n),\\
&\overline{\mathfrak{p}_3}(3^n)=\textrm{span}\{\overline{D}\otimes A|A\in \mathfrak{u}(3^{n-1})\},\\
&\overline{\mathfrak{a}_3}(3^n)=\textrm{span}\{\overline{D}\otimes A|A\in\overline{\alpha}(3^{n-1})\}.\\
\end{aligned}
\end{eqnarray}
Here $\overline{\mathfrak{a}_3}(3^n)$ is the Cartan subalgebra of $(\overline{\mathfrak{l}_2}(3^n),\mathfrak{l}_3(3^n))$ and
\begin{eqnarray}\label{eq83}
\begin{aligned}
&\overline{\alpha}(3)=\{\textrm{i}I_{3},\textrm{i}\overline{D},\textrm{i}\sigma_{z}^{01}\},\\
&\overline{\alpha}(3^n)=\{I_{3}\otimes A,\overline{D}\otimes A,\sigma_{z}^{01}\otimes A|A\in \overline{\alpha}(3^{n-1})\}.
\end{aligned}
\end{eqnarray}

\textbf{Proof 6.} Given any $A_1, A_2\in\mathfrak{u}(3^{n-1})$, the following relations
\begin{eqnarray}\label{eq84}
\begin{aligned}
&[I_3\otimes A_1,D\otimes A_2]=D\otimes [A_1,A_2],\\
&[D\otimes A_1,D\otimes A_2]=I_3\otimes [A_1,A_2],\\
&[I_3\otimes A_1,\overline{D}\otimes A_2]=\overline{D}\otimes [A_1,A_2],\\
&[\overline{D}\otimes A_1,\overline{D}\otimes A_2]=I_3\otimes [A_1,A_2],
\end{aligned}
\end{eqnarray}
indicate that $\mathfrak{l}_3(3^n)$ and $\mathfrak{p}_3(3^n)$ ($\overline{\mathfrak{p}_3}(3^n)$) satisfy Eq. (\ref{eq2}).
From Eq. (\ref{eq17}), it is obvious that
\begin{eqnarray}\label{eq85}
\textrm{span}\{\overline{\alpha}(3^n)\}=\textrm{span}\{\widetilde{\alpha}(3^n)\}=\textrm{span}\{\alpha(3^n)\},
\end{eqnarray}
By Lemme 1, $\mathfrak{a}_3(n)$ and $\overline{\mathfrak{a}_3}(n)$ are Cartan subalgebras.
$\hfill\blacksquare$

\begin{figure} [htbp]
  \centering
  \subfigure[]{
  \includegraphics[width=5cm]{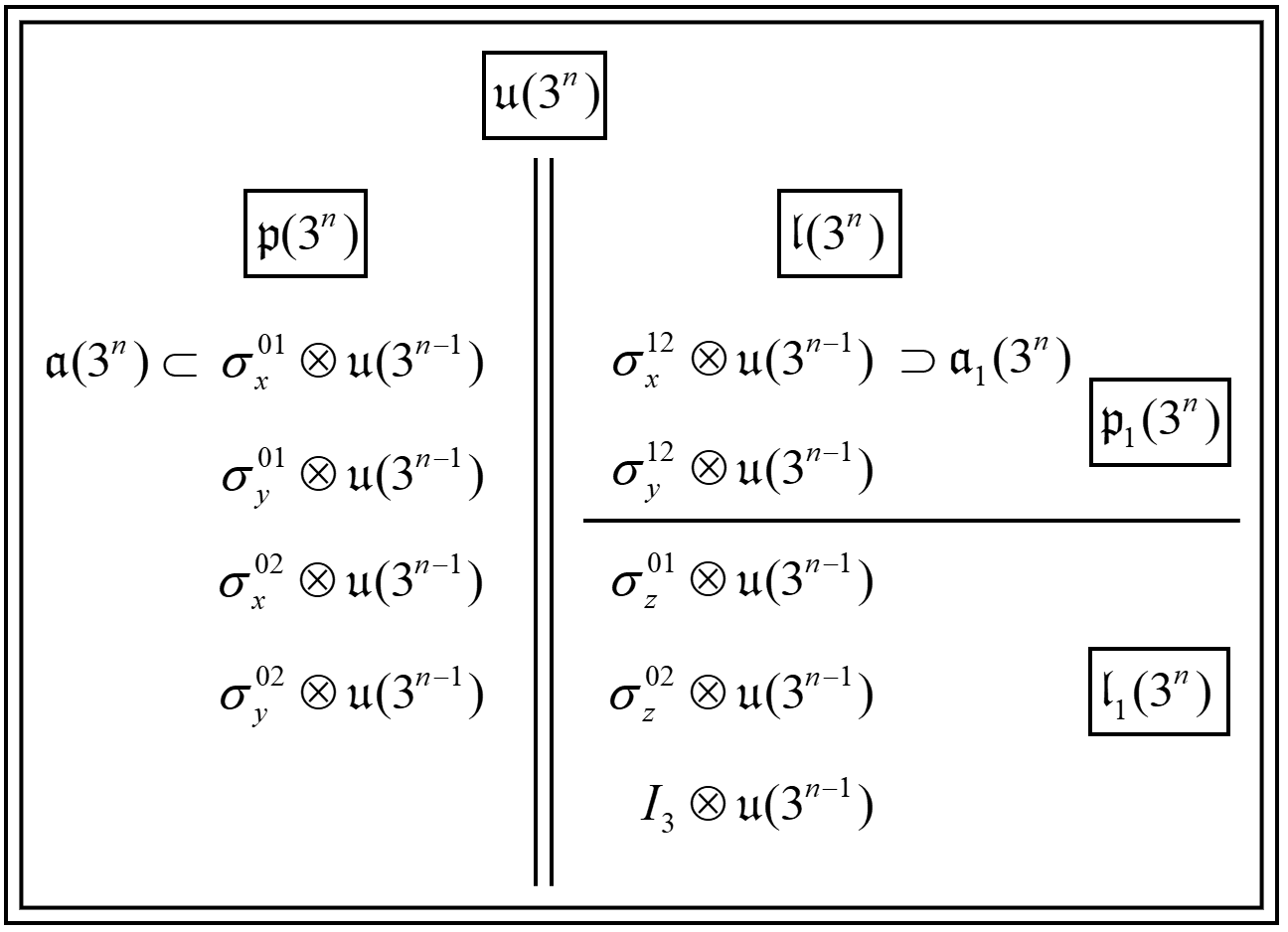}
 \label{9a} }
  \subfigure[]{
  \includegraphics[width=4.05cm]{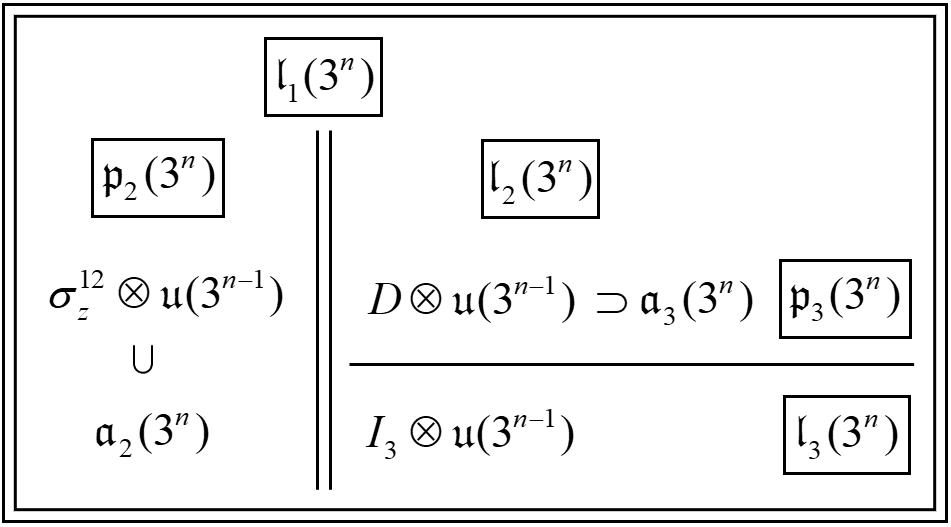}
  \label{9b}}
  \subfigure[]{
  \includegraphics[width=4.05cm]{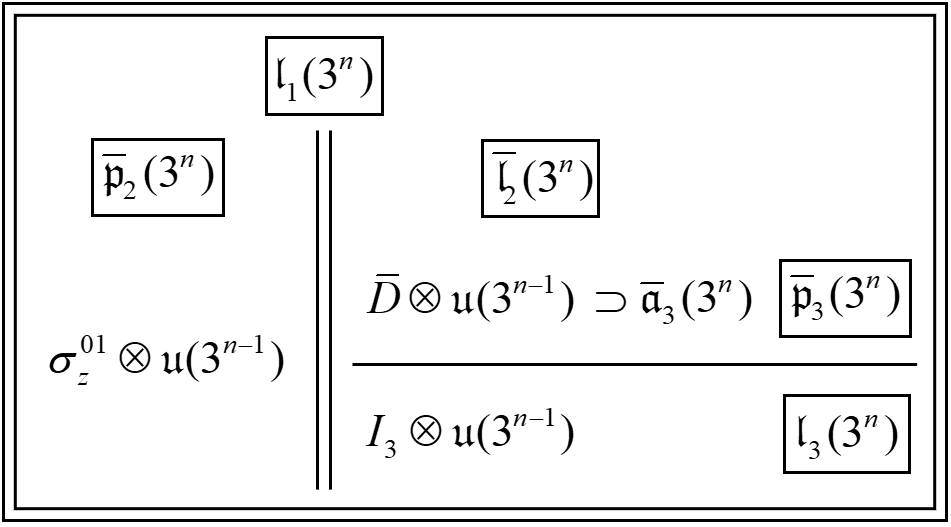}
  \label{9c}}
  \caption{(a) Cartan decomposition of $\mathfrak{u}(3^n)$ and $\mathfrak{l}(3^n)$.
  (b) Cartan decomposition of $\mathfrak{l}_1(3^n)$ and $\mathfrak{l}_2(3^n)$ with the diagonal basis $\widetilde{\alpha}(3^n)$. (c) Cartan decomposition of $\mathfrak{l}_1(3^n)$ and $\overline{\mathfrak{l}_2}(3^n)$ with the diagonal basis $\overline{\alpha}(3^n)$.}
  \label{Fig.9}
\end{figure}

Based on the above successive Cartan decompositions of $\mathfrak{u}(3^n)$ depicted by Fig. \ref{Fig.9}, we can reduce the $n$-qutrit gate to local and nonlocal operations, see Corollary 2.
Its proof is a direct extension of that of Corollary 1.

\textbf{Corollary 2.} Any $M\in U(3^n)$ can be decomposed as
\begin{eqnarray}\label{eq86}
\begin{aligned}
M=&K_1 \cdot \overline{A}^{(3)}_1 \cdot K_2 \cdot A^{(1)}_1 \cdot K_3 \cdot A^{(3)}_1 \cdot K_4 \cdot A \cdot K_5\\ &\cdot \overline{A}^{(3)}_2 \cdot K_6 \cdot A^{(1)}_2 \cdot K_7 \cdot A^{(2)} \cdot K_8 \cdot A^{(3)}_2\cdot K_9.
\end{aligned}
\end{eqnarray}
Here $A\in \textrm{exp}[\mathfrak{a}(3^n)]$, $A^{(1)}_{1,2}\in \textrm{exp}[\mathfrak{a}_1(3^n)]$, $A^{(2)}\in \textrm{exp}[\mathfrak{a}_2(3^n)]$, $A^{(3)}_{1,2}\in \textrm{exp}[\mathfrak{a}_3(3^n)]$,
$\overline{A}^{(3)}_{1,2}\in \textrm{exp}[\overline{\mathfrak{a}_3}(3^n)]$, and $K_i\in \textrm{exp}[\mathfrak{l}_3(3^n)]$ is an ($n-1$)-qutrit gate.

\subsection{Synthesis of arbitrary \emph{n}-qutrit gates}\label{sec4.2}

Note that $K_i$ can be decomposed again into the form of Eq. (\ref{eq86}) as $\mathfrak{l}_3(3^n)$ is isomorphic to $\mathfrak{u}(3^{n-1})$.
To complete the construction of arbitrary $n$-qutrit gates in terms of single-qutrit type rotations and GCXs, next we synthesize nonlocal operations $A$, $A^{(1)}_{1,2}$, $A^{(2)}$, $A^{(3)}_{1,2}$, and $\overline{A}^{(3)}_{1,2}$.

\textbf{Synthesis of \emph{n}-qutrit operation \bm{$A$}}.
Based on Eq. (\ref{eq47}), $n$-qutrit nonlocal operation $A$ has the form
\begin{eqnarray}\label{eq87}
\begin{aligned}
A(\hat{\theta})=\textrm{exp}[-\textrm{i}\sigma_x^{01} \otimes A_1(\hat{\theta})].
\end{aligned}
\end{eqnarray}
Here $\textrm{i}A_1(\hat{\theta})\in \text{span} \{\alpha(3^{n-1})\}$ is a $3^{n-1}\times 3^{n-1}$ diagonal matrix depending on $3^{n-1}$ real parameters. Hence without loss of generality $A_1(\hat{\theta})$ has the form
\begin{eqnarray}\label{eq88}
\begin{aligned}
A_1(\hat{\theta})=\textrm{diag}\{&\underline{\theta_{11},\theta_{12},\theta_{13}},
                                  \underline{\theta_{21},\theta_{22},\theta_{23}},
                                  \underline{\theta_{31},\theta_{32},\theta_{33}},
\\&\cdots,\underline{\theta_{3^{n-2}1},\theta_{3^{n-2}2},\theta_{3^{n-2}3}}\}.
\end{aligned}
\end{eqnarray}

Using Theorem 1, we obtain
\begin{eqnarray}\label{eq89}
\begin{aligned}
A(\hat{\theta})=&R_y^{01}(\frac{\pi}{2})\otimes I_{3^{n-1}}\cdot\textrm{exp}[-\textrm{i}\sigma_z^{01} \otimes A_1(\hat{\theta})]\\&
\cdot R_y^{01}(-\frac{\pi}{2})\otimes I_{3^{n-1}}\\
=&R_y^{01}(\frac{\pi}{2})\otimes I_{3^{n-1}} \cdot \textrm{GCX}_n^1(1\rightarrow X^{01})\\&
\cdot\textrm{exp}[-\textrm{i}\sigma_z^{01} \otimes A_2(\hat{\theta}_1)]\otimes I_3\\&
\cdot\textrm{GCX}_n^1(0\rightarrow X^{01})\cdot X^{01}\otimes I_{3^{n-1}}\\&
\cdot\textrm{exp}[-\textrm{i}\sigma_z^{01} \otimes A_2(\hat{\theta}_2)]\otimes I_3\\&
\cdot\textrm{GCX}_n^1(2\rightarrow X^{01})\\&
\cdot\textrm{exp}[-\textrm{i}\sigma_z^{01} \otimes A_2(\hat{\theta}_3)]\otimes I_3\\&
\cdot R_y^{01}(-\frac{\pi}{2})\otimes I_{3^{n-1}},
\end{aligned}
\end{eqnarray}
where
\begin{eqnarray}\label{eq90}
\begin{aligned}
A_2(\hat{\theta}_1)=\textrm{diag}\{&\frac{\theta_{11}-\theta_{12}}{2},\frac{\theta_{21}-\theta_{22}}{2},\cdots,\\&\frac{\theta_{3^{n-2}1}-\hat{\theta}_{3^{n-2}2}}{2}\},\\
A_2(\hat{\theta}_2)=\textrm{diag}\{&\frac{\theta_{11}-\theta_{13}}{2},\frac{\theta_{21}-\theta_{23}}{2},\cdots,\\&\frac{\theta_{3^{n-2}1}-\hat{\theta}_{3^{n-2}3}}{2}\},\\
A_2(\hat{\theta}_3)=\textrm{diag}\{&\frac{\theta_{12}+\theta_{13}}{2},\frac{\theta_{22}+\theta_{23}}{2},\cdots,\\&\frac{\theta_{3^{n-2}2}+\hat{\theta}_{3^{n-2}3}}{2}\}.
\end{aligned}
\end{eqnarray}

\begin{figure} [htbp]
  \centering
  \includegraphics[width=8.3cm]{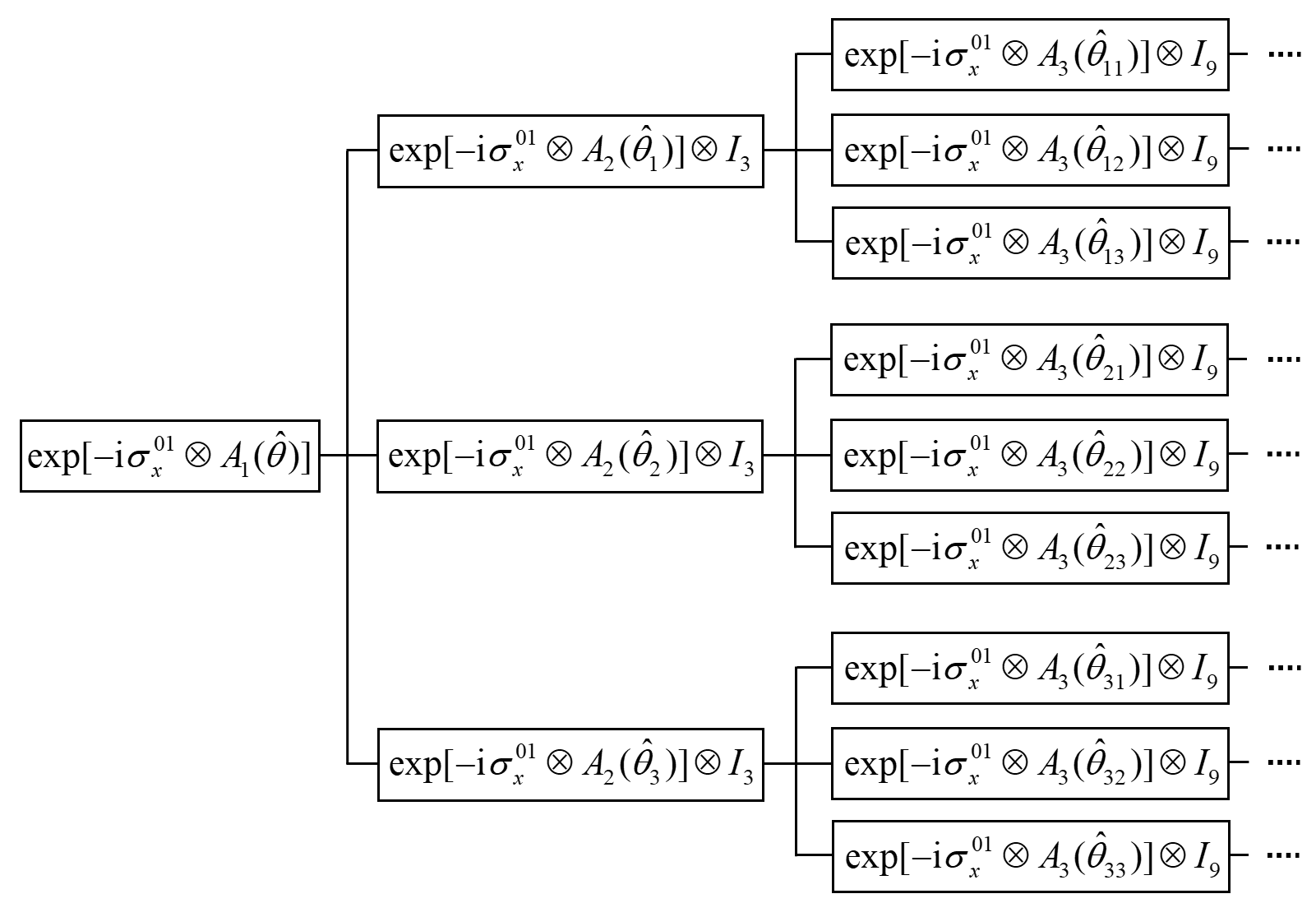}
  \caption{The recursive decomposition of $n$-qutrit gate $A(\hat{\theta})$. Here $\textrm{i}A_k\in \text{span}\{\alpha(3^{n-k})\}$ and $\textrm{exp}[-\textrm{i}\sigma_z^{01} \otimes A_{k}]$ is an $(n-k+1)$-qutrit gate.}
  \label{Fig.10}
\end{figure}

It is remarkable that $\textrm{i}A_2(\hat{\theta}_i)\in \textrm{span}\{\alpha(3^{n-2})\}$.
Thus, Eq. (\ref{eq89}) is a recursive decomposition until $\textrm{i}A_{n-1}\in \textrm{span}\{\alpha(3)\}$, see Fig. \ref{Fig.10}.

\textbf{Synthesis of \emph{n}-qutrit operation \bm{$A^{(1)}_{1,2}$}}.
A similar argument as that made in $A$, the iterative decomposition of $A_{1,2}^{(1)}$ is given by
\begin{eqnarray}\label{eq91}
\begin{aligned}
A^{(1)}(\hat{\theta})=&\textrm{exp}[-\textrm{i}
\sigma_x^{12} \otimes A_1(\hat{\theta})]\\
=&R_y^{12}(\frac{\pi}{2})\otimes I_{3^{n-1}} \cdot \textrm{GCX}_n^1(1\rightarrow X^{12})\\&
\cdot\textrm{exp}[-\textrm{i}\sigma_z^{12} \otimes A_2(\hat{\theta}_1)]\otimes I_3\\&
\cdot\textrm{GCX}_n^1(0\rightarrow X^{12})\cdot X^{12}\otimes I_{3^{n-1}}\\&
\cdot\textrm{exp}[-\textrm{i}\sigma_z^{12} \otimes A_2(\hat{\theta}_2)]\otimes I_3\\&
\cdot\textrm{GCX}_n^1(2\rightarrow X^{12})\\&
\cdot\textrm{exp}[-\textrm{i}\sigma_z^{12} \otimes A_2(\hat{\theta}_3)]\otimes I_3\\&
\cdot R_y^{12}(-\frac{\pi}{2})\otimes I_{3^{n-1}}.
\end{aligned}
\end{eqnarray}

\begin{figure} [htbp]
  \centering
  \includegraphics[width=8.5cm]{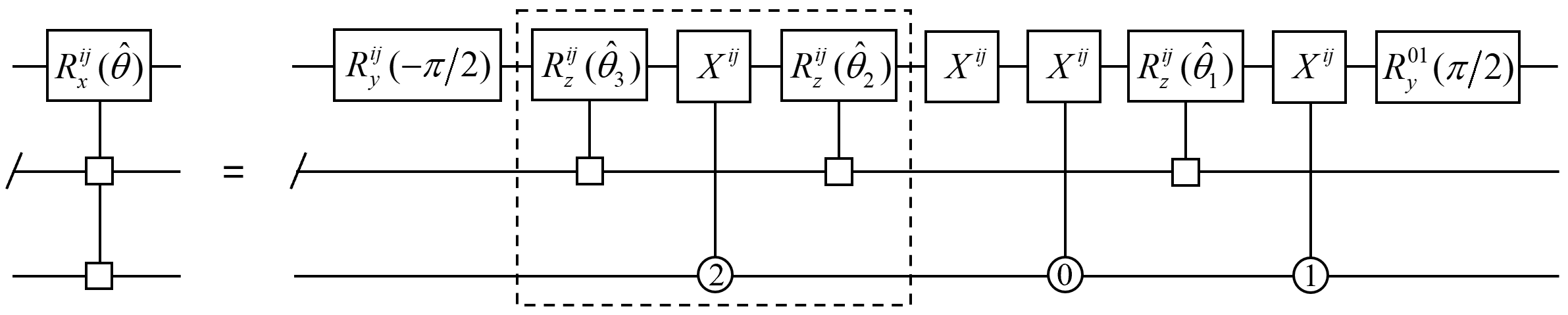}
  \caption{Synthesis of $n$-qutrit gate $A$ with $ij=01$ or $A^{(1)}_{1,2}$ with $ij=12$. Note that $X^{ij}$ can be decomposed into three single-qutrit type rotation gates by Eq. (\ref{eq31}) or (\ref{eq34}) up to a global phase $e^{\textrm{i}\frac{\pi}{3}}$.}
  \label{Fig.11}
\end{figure}

An $n$-qutrit gate with matrix form $\textrm{exp}[-\textrm{i}\sigma_{\varphi}^{ij} \otimes A_1(\hat{\theta})]$ or $\textrm{exp}[-\textrm{i} A_1(\hat{\theta})\otimes\sigma_{\varphi}^{ij}]$ is a three-valued uniformly $(n-1)$-fold controlled $R_{\varphi}^{ij}$ rotation in Ref. \cite{Di2013}.
In quantum circuit diagrams, the square ($\Box$) denotes the control qutrits of such logic gate and the slash ($/$) denotes multiple qutrits on the line.
Following this manner, the synthesized circuit of $A(\hat{\theta})$ and $A^{(1)}(\hat{\theta})$ is presented in Fig. \ref{Fig.11}.

\begin{figure} [htbp]
  \centering
  \includegraphics[width=7cm]{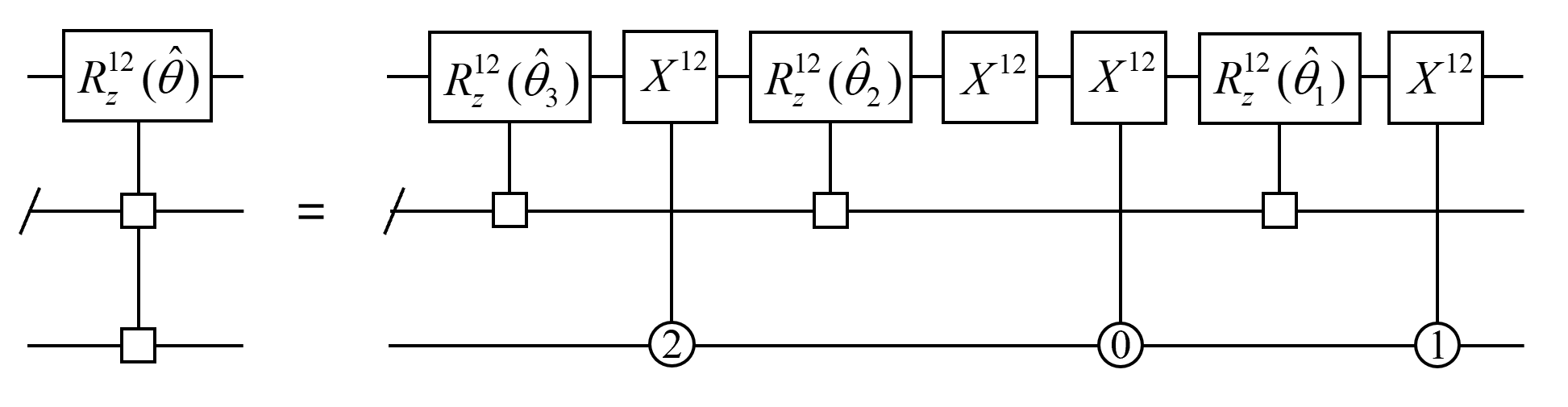}
  \caption{Synthesis of $n$-qutrit gate $A^{(2)}$.}
  \label{Fig.12}
\end{figure}

\textbf{Synthesis of \emph{n}-qutrit operation \bm{$A^{(2)}$}}.
Since the relation span$\{\widetilde{\alpha}(3^n)\}=\text{span}\{\alpha(3^n)\}$, we can obtain
\begin{eqnarray}\label{eq92}
\begin{aligned}
A^{(2)}(\hat{\theta})=&\textrm{exp}[-\textrm{i}\sigma_z^{12} \otimes A_1(\hat{\theta})]\\
=&R_y^{12}(-\frac{\pi}{2})\otimes I_{3^{n-1}}\cdot\textrm{exp}[-\textrm{i}\sigma_x^{12} \otimes A_1(\hat{\theta})]\\&\cdot R_y^{12}(\frac{\pi}{2})\otimes I_{3^{n-1}}.
\end{aligned}
\end{eqnarray}
Based on Eq. (\ref{eq91}), one can get that the recursive decomposition of $A^{(2)}(\hat{\theta})$ as follows
\begin{eqnarray}\label{eq93}
\begin{aligned}
A^{(2)}(\hat{\theta})=& \textrm{GCX}_n^1(1\rightarrow X^{12})\\&
\cdot\textrm{exp}[-\textrm{i}\sigma_z^{12} \otimes A_2(\hat{\theta}_1)]\otimes I_3\\&
\cdot\textrm{GCX}_n^1(0\rightarrow X^{12})\cdot X^{12}\otimes I_{3^{n-1}}\\&
\cdot\textrm{exp}[-\textrm{i}\sigma_z^{12} \otimes A_2(\hat{\theta}_2)]\otimes I_3\\&
\cdot\textrm{GCX}_n^1(2\rightarrow X^{12})\\&
\cdot\textrm{exp}[-\textrm{i}\sigma_z^{12} \otimes A_2(\hat{\theta}_3)]\otimes I_3.
\end{aligned}
\end{eqnarray}
The synthesis of $A^{(2)}$ is given by Fig. \ref{Fig.12}.

\textbf{Synthesis of \emph{n}-qutrit operation \bm{$A_{1,2}^{(3)}$} and \bm{$\overline{A}^{(3)}_{1,2}$}}.
Based on Eq. (\ref{eq81}), $A^{(3)}_{1,2}$ can be given by
\begin{eqnarray}\label{eq94}
\begin{aligned}
A^{(3)}(\hat{\theta})=&\textrm{exp}[-\textrm{i}D \otimes A_1(\hat{\theta})]\\
=&\textrm{GCX}_1^n(0\rightarrow X^{02})\\&
\cdot I_3\otimes \textrm{exp}[-\textrm{i}A_2(\hat{\theta}_4)\otimes \sigma_z^{02}]\\&
\cdot \textrm{GCX}_1^n(0\rightarrow X^{02})\\&
\cdot\textrm{GCX}_1^n(0\rightarrow X^{01})\\&
\cdot I_3\otimes \textrm{exp}[-\textrm{i}A_2(\hat{\theta}_5)\otimes \sigma_z^{01}]\\&
\cdot \textrm{GCX}_1^n(0\rightarrow X^{01})\\&
\cdot \textrm{exp}[-\textrm{i}D\otimes A_2(\hat{\theta}_6)]\otimes I_3,
\end{aligned}
\end{eqnarray}
where
\begin{eqnarray}\label{eq95}
\begin{aligned}
A_2(\hat{\theta}_4)=\textrm{diag}\{&\frac{2\theta_{13}-\theta_{11}-\theta_{12}}{3},\frac{2\theta_{23}-\theta_{21}-\theta_{22}}{3},\\&\cdots,\frac{2\theta_{3^{n-2}3}-\theta_{3^{n-2}1}-\theta_{3^{n-2}2}}{3}\},\\
A_2(\hat{\theta}_5)=\textrm{diag}\{&\frac{2\theta_{12}-\theta_{11}-\theta_{13}}{3},\frac{2\theta_{22}-\theta_{21}-\theta_{23}}{3},\\&\cdots,\frac{2\theta_{3^{n-2}2}-\theta_{3^{n-2}1}-\theta_{3^{n-2}3}}{3}\},\\
A_2(\hat{\theta}_6)=\textrm{diag}\{&\frac{\theta_{11}+\theta_{12}+\theta_{13}}{3},\frac{\theta_{21}+\theta_{22}+\theta_{23}}{3},\\&\cdots,\frac{\theta_{3^{n-2}1}+\theta_{3^{n-2}2}+\theta_{3^{n-2}3}}{3}\}.
\end{aligned}
\end{eqnarray}
Observe that Eq. (\ref{eq94}) can be applied to recursively decompose the last term $\textrm{exp}[-\textrm{i}D\otimes A_2(\hat{\theta}_6)]$.

Since span$\{\overline{\alpha}(3^n)\}=\text{span}\{\alpha(3^n)\}$, $\overline{A}^{(3)}_{1,2}$ has a decomposition form similar to $A^{(3)}_{1,2}$ as follows
\begin{eqnarray}\label{eq96}
\begin{aligned}
\overline{A}^{(3)}(\hat{\theta})=&\textrm{exp}[-\textrm{i}\overline{D} \otimes A_1(\hat{\theta})]\\
=&\textrm{GCX}_1^n(2\rightarrow X^{02})\\&
\cdot I_3\otimes \textrm{exp}[-\textrm{i}A_2(\hat{\theta}_4)\otimes \sigma_z^{02}]\\&
\cdot \textrm{GCX}_1^n(2\rightarrow X^{02})\\&
\cdot\textrm{GCX}_1^n(2\rightarrow X^{01})\\&
\cdot I_3\otimes \textrm{exp}[-\textrm{i}A_2(\hat{\theta}_5)\otimes \sigma_z^{01}]\\&
\cdot \textrm{GCX}_1^n(2\rightarrow X^{01})\\&
\cdot \textrm{exp}[-\textrm{i}\overline{D}\otimes A_2(\hat{\theta}_6)]\otimes I_3,
\end{aligned}
\end{eqnarray}

\begin{figure} [htbp]
  \centering
  \includegraphics[width=7cm]{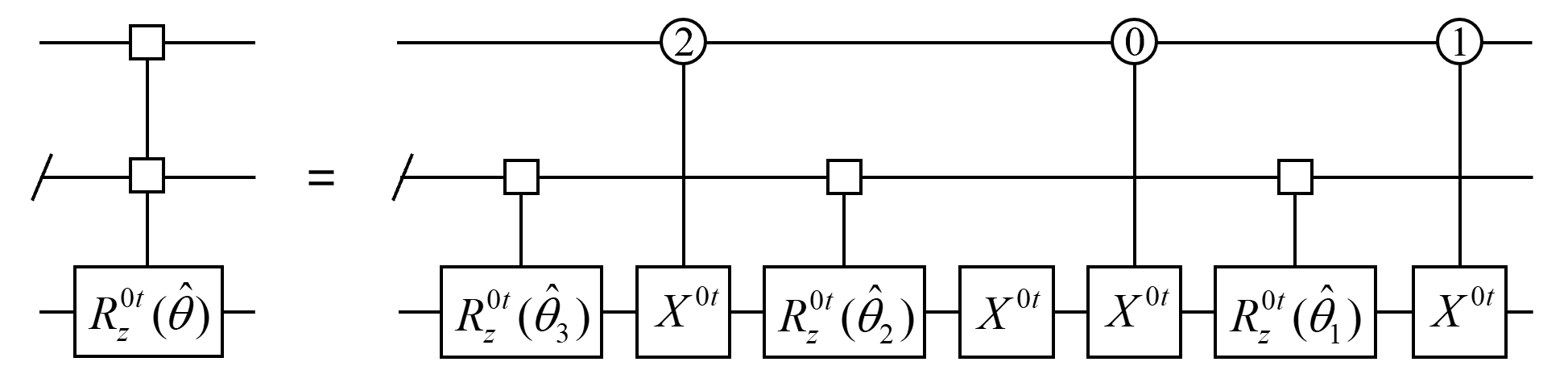}
  \caption{Synthesis of $n$-qutrit gate $\textrm{exp}[-\textrm{i}A_1(\hat{\theta})\otimes \sigma_z^{0t}]$ $(t=1,2)$.}
  \label{Fig.13}
\end{figure}

\begin{figure} [htbp]
  \centering
  \includegraphics[width=6cm]{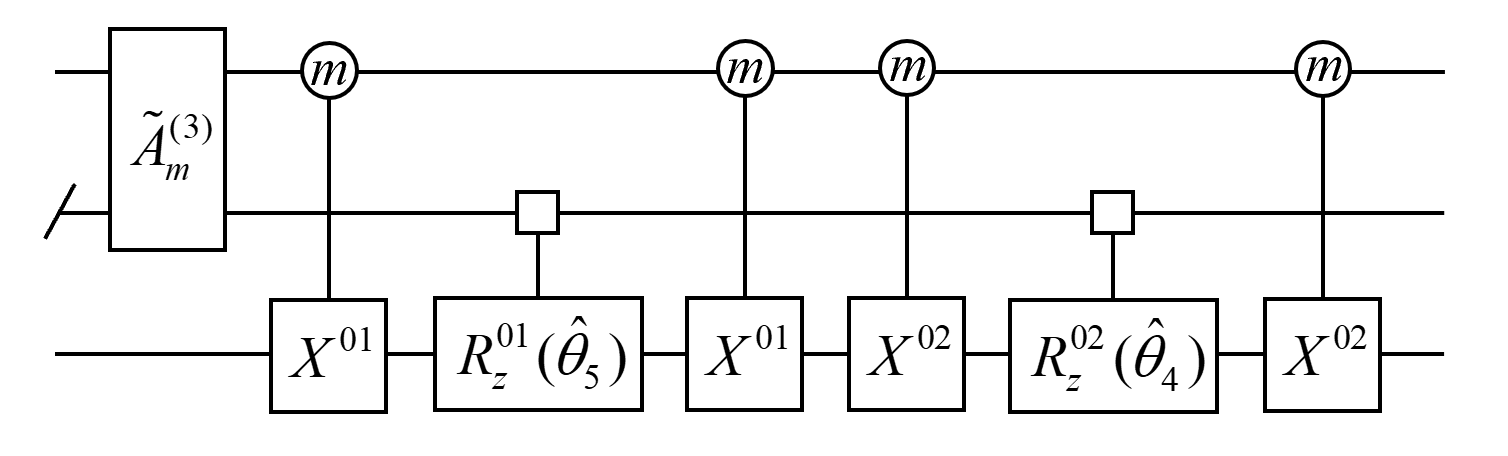}
  \caption{Synthesis of $n$-qutrit gate $A_{1,2}^{(3)}$ with $m=0$ or $\overline{A}_{1,2}^{(3)}$ with $m=2$. Here the $(n-1)$-qutrit gates $\widetilde{A}_0^{(3)}=\textrm{exp}[-\textrm{i}D\otimes A_2(\hat{\theta}_6)]$ and $\widetilde{A}_2^{(3)}=\textrm{exp}[-\textrm{i}\overline{D}\otimes A_2(\hat{\theta}_6)]$. Note that two adjacent GCX gates here can be combined into a CINC gate.}
  \label{Fig.14}
\end{figure}

In addition, the synthetic circuit of the $n$-qutrit gate $\textrm{exp}[-\textrm{i}A_1(\hat{\theta})\otimes \sigma_z^{0t}]$ $(t=1,2)$ is shown in Fig. \ref{Fig.13}. Thus, we have completed the recursive decompositions of $A^{(3)}_{1,2}(\hat{\theta})$ and $\overline{A}^{(3)}_{1,2}(\hat{\theta})$ and Fig. \ref{Fig.14} presents the first step of the decomposition.

\textbf{Synthesis of the arbitrary \emph{n}-qutrit gate}. 
By successively applying Corollary 2 to decompose $K_i$, $M$ can be decomposed into generic two-qutrit gates and specific non-local operations $A$, $A^{(1)}$, $A^{(2)}$, $A^{(3)}$, and $\overline{A}^{(3)}$.
The synthesis of components as two-qutrit gates is presented in Sec. \ref{sec3}, and then in conjunction with Fig. \ref{Fig.11}, Fig. \ref{Fig.12}, Fig. \ref{Fig.13}, and Fig. \ref{Fig.14}, the synthesis of arbitrary $n$-qutrit gates is accomplished.

Similarly to the case of $n=2$, the rightmost GCX of Fig. \ref{Fig.11} can be absorbed.
Since $\textrm{GCX}^k_{l_1}(m_1\rightarrow X^{ij})$ and $\textrm{GCX}^k_{l_2}(m_2\rightarrow X^{ij})$ are commutative, the structure shown in the dotted box of Fig. \ref{Fig.11} has two GCX gates that are cancelled out in the next step of the decomposition.
Based on the above discussion, the required GCX gates for these non-local operators are shown in Table \ref{Table1}.

\begin{table}[htbp]
\centering
\caption{The GCX gate count for the synthesis of $n$-qutrit gates $A$, $A^{(1)}_{1,2}$, $A^{(2)}$, $A^{(3)}_{1,2}$, and $\overline{A}^{(3)}_{1,2}$. }\label{Table1}
\begin{tabular}{ccccc}
\hline  \hline
  & 2  & 3  & 4  & $n\geq2$  \\
\hline
\makebox[0.1\textwidth]{$A/A^{(1)}_{1,2}$} & \makebox[0.045\textwidth]{2} & \makebox[0.045\textwidth]{8} & \makebox[0.045\textwidth]{26} & \makebox[0.1\textwidth]{$3^{n-1}-1$} \\ 
$A^{(2)}$ & 3 & 10 & 29 & $3^{n-1}+n-2$ \\ 
$A^{(3)}_{1,2}/\overline{A}^{(3)}_{1,2}$ & 4 & 14 & 38 & $3^{n-1}+n^2-n-1$ \\ 
\hline  \hline
\end{tabular}
\end{table}

According to Table \ref{Table1}, the GCX gate count for an exact synthesis of $M$ can be calculated as
\begin{eqnarray}\label{eq97}
\begin{aligned}
\frac{47}{96}\cdot3^{2n}-4\cdot3^{n-1}-(\frac{n^2}{2}+\frac{3n}{4}-\frac{27}{32}).
\end{aligned}
\end{eqnarray}

\section{Conclusion}\label{sec5}

Multi-valued quantum computing is currently a well-attended research area, as the advantages of qudit circuits are gradually being discovered.
In this paper, we propose a Cartan decomposition of $U(3^2)$ that divides arbitrary two-qutrit gates into single-qutrit gates and non-local transitions, i.e., $A$, $A^{(1)}$, $A^{(2)}$, $A^{(3)}$, and $\overline{A}^{(3)}$.
Based on Corollary 1 derived from the decomposition, we design a quantum circuit of generic two-qutrit gates as shown in Fig. \ref{Fig.3}, where the non-local operations are simulated exactly by simple elementary gates, i.e., GCX, CINC, and $R_\varphi^{ij}(\theta)$ gates.
In the presented two-qutrit circuit, up to 21 GCXs and CINCs are necessary, whereas the previously optimal circuit required 26 in the worst case \cite{Di2015}.

Furthermore, a recursive Cartan decomposition of $U(3^n)$ and its proof are given, which divides arbitrary $n$-qutrit gates into $(n-1)$-qutrit gates that can be decomposed again and non-local transformations described in Corollary 2.
Relying on Corollary 2, we propose a recursive algorithm to synthesize an arbitrary $n$-qutrit gate that terminates in the two-qutrit case.
The quantum circuit of generic $n$-qutrit gates constructed by the recursive program is asymptotically optimal.
When only GCX gates are chosen as the elementary two-qutrit gate to simulate the non-local operators, the total number of GCX gates required for the synthetic circuit is shown as Eq. (\ref{eq97}), which is less than the previously optimal QSD-based circuit \cite{Di2013,Di2015}.
Remarkably, when the CINC gates have been introduced as elementary gates, the cost of the presented circuit can be further reduced by $\frac{1}{16}\cdot3^{2n}-\frac{n}{2}-\frac{1}{16}$ elementary two-qutrit gates.
As shown in Table \ref{Table2}, we have compared it with the previous QR-, CSD-, and QSD-based synthesis, which concludes our program is the optimum.

\begin{table}[htbp]
\centering
\caption{The comparison of elementary two-qutrit gate count for the synthesis of a general $n$-qutrit gate.}\label{Table2}
\begin{tabular}{ccccc}
\hline  \hline
  & \makebox[0.035\textwidth]{2} &  \makebox[0.05\textwidth]{3} & 4  & \makebox[0.15\textwidth]{$n\geq2$}  \\
\hline
QR \cite{Bullock2005} & 45 & 675 & 6561  & $3^{2n}+3^n\cdot(2n-8)$ \\
CSD \cite{Nakajima2009}& 36 & 3360 & 20088  &—— \\
QSD \cite{Di2013}& 44 & 692 & 6860  &—— \\
QSD,optimal \cite{Di2015}& 26 & 344 & 3458  &$O(\alpha3^{2n})$\\  \cline{2-5}
\multirow{2}{*}{Our work} & \multirow{2}{*}{21} & \multirow{2}{*}{217} & \multirow{2}{*}{2686}  & $\frac{41}{96}\cdot3^{2n}-4\cdot3^{n-1}$ \\
&  & & & $-(\frac{n^2}{2}+\frac{n}{4}-\frac{29}{32})$ \\
\hline  \hline
\end{tabular}
\end{table}

With the improvement of optical implementations of three-valued quantum systems, the proposed synthesis facilitates the physical implementation of $n$-qutrit gates.
In addition, to coincide with the lower bound of complexity for generic $n$-qutrit circuits, there may be room for improvement through optimizing non-local transformations and reducing $(n-1)$-qutrit gates generated by the first step of the decomposition.

\section*{ACKNOWLEDGEMENTS} \par

This work is supported by the National Natural Science Foundation of China under Grant No. 62371038 and the Fundamental Research Funds for the Central Universities under Grant No. FRF-TP-19-011A3.

\section*{appendix}

\appendix

\section{Proof of Corollary 1}\label{Appendix.A}

Based on the Cartan decomposition of $\mathfrak{u}(9)$ and $\mathfrak{l}(9)$ as shown Fig. \ref{2a}, for any $M\in U(9)$, we have
\begin{eqnarray}\label{A1}
\begin{aligned}
M=K_1' \cdot A^{(1)}_1 \cdot K_2' \cdot A \cdot K_3' \cdot A^{(1)}_2 \cdot K_4'.
\end{aligned}
\end{eqnarray}
Here $K_i'\in \textrm{exp}[\mathfrak{l}_1(9)]$, $A\in \textrm{exp}[\mathfrak{a}(9)]$, and $A^{(1)}_{1,2}\in \textrm{exp}[\mathfrak{a}_1(9)]$.
From the definition of $\mathfrak{l}_1(9)$, $K_i'$ can be written as
\begin{eqnarray}\label{A2}
\begin{aligned}
K_i'
=\left(
  \begin{array}{ccc}
    U_{i1} & \textbf{0} & \textbf{0}\\
    \textbf{0} & U_{i2} & \textbf{0}\\
    \textbf{0} & \textbf{0} & U_{i3}\\
  \end{array}
\right),
\end{aligned}
\end{eqnarray}
where $U_{ij}\in U(3)$ and $i=1,2,3,4$, $j=1,2,3$.
According to the subalgebras $\mathfrak{a}(9)$ and $\mathfrak{a}_1(9)$ chosen as in Eqs. (\ref{eq12}) and (\ref{eq14}), $A$ and $A^{(1)}_{1,2}$ can be expressed as
\begin{eqnarray}\label{A3}
\begin{aligned}
A
&=\left(
  \begin{array}{ccc}
    C_0 & -\textrm{i}S_0 & \textbf{0}\\
    -\textrm{i}S_0 & C_0 & \textbf{0}\\
    \textbf{0} & \textbf{0} & I_3\\
  \end{array}
\right),\\
A^{(2)}_1
&=\left(
  \begin{array}{ccc}
    I_3 & \textbf{0} & \textbf{0}\\
    \textbf{0} & C_1 & -\textrm{i}S_1\\
    \textbf{0} & -\textrm{i}S_1 & C_1\\
  \end{array}
\right),\\
A^{(2)}_2
&=\left(
  \begin{array}{ccc}
    I_3 & \textbf{0} & \textbf{0}\\
    \textbf{0} & C_2 & -\textrm{i}S_2\\
    \textbf{0} & -\textrm{i}S_2 & C_2\\
  \end{array}
\right).
\end{aligned}
\end{eqnarray}
Here $C_k=\textrm{diag}\{\textrm{cos}\alpha_{k1},\textrm{cos}\alpha_{k2},\textrm{cos}\alpha_{k3}\}$ and
$S_k=\textrm{diag}\{\textrm{sin}\alpha_{k1},\textrm{sin}\alpha_{k2},\textrm{sin}\alpha_{k3}\}$, and $k=0,1,2$.

Substituting Eqs. (\ref{A2}) and (\ref{A3}) into Eq. (\ref{A1}) and using the multiplication of the block diagonal matrices, we can get
\begin{eqnarray}\label{A4}
\begin{aligned}
M=K_1''\cdot A^{(1)}_1 \cdot K_2''\cdot A \cdot K_3''\cdot A^{(1)}_2 \cdot K_4''.
\end{aligned}
\end{eqnarray}
where
\begin{eqnarray}\label{A5}
\begin{aligned}
&K_1''=\left(
  \begin{array}{ccc}
    U_{12} & \textbf{0} & \textbf{0}\\
    \textbf{0} & U_{12} & \textbf{0}\\
    \textbf{0} & \textbf{0} & U_{13}\\
  \end{array}
\right),\\
&K_2''=\left(
  \begin{array}{ccc}
    U_{12}^{\dag}U_{11}U_{21} & \textbf{0} & \textbf{0}\\
    \textbf{0} & U_{22} & \textbf{0}\\
    \textbf{0} & \textbf{0} & U_{22}\\
  \end{array}
\right),\\
&K_3''=\left(
  \begin{array}{ccc}
    U_{32} & \textbf{0} & \textbf{0}\\
    \textbf{0} & U_{32} & \textbf{0}\\
    \textbf{0} & \textbf{0} & U_{22}^{\dag}U_{23}U_{33}\\
  \end{array}
\right),\\
&K_4''=\left(
  \begin{array}{ccc}
    U_{32}^{\dag}U_{31}U_{41} & \textbf{0} & \textbf{0}\\
    \textbf{0} & U_{42} & \textbf{0}\\
    \textbf{0} & \textbf{0} & U_{43}\\
  \end{array}
\right).
\end{aligned}
\end{eqnarray}
Observe that $K_{1,3}''\in\textrm{exp}[\overline{\mathfrak{l}_2}(9)$], $K_{2}''\in\textrm{exp}[\mathfrak{l}_2(9)]$, and $K_{4}''\in\textrm{exp}[\mathfrak{l}_1(9)]$.
Based on the Cartan decomposition of $\mathfrak{l}_2(9)$ and $\overline{\mathfrak{l}_2}(9)$ given by Eqs. (\ref{eq23}) and (\ref{eq25}), we then have
\begin{eqnarray}\label{A6}
\begin{aligned}
&K_1''=K_1\cdot \overline{A}^{(3)}_1 \cdot K_2,\\
&K_2''=K_3\cdot A^{(3)}_1 \cdot K_4,\\
&K_3''=K_5\cdot \overline{A}^{(3)}_2 \cdot K_6,\\
\end{aligned}
\end{eqnarray}
where $A^{(3)}_{1}\in \textrm{exp}[\mathfrak{a}_3(9)]$,
$\overline{A}^{(3)}_{1,2}\in \textrm{exp}[\overline{\mathfrak{a}_3}(9)]$, and $K_r\in \textrm{exp}[\mathfrak{l}_3(9)]$, $r=1,2,\cdots,6$.

As for $K_4''$, the decomposition of $\mathfrak{l}_1(9)$ shown in Eq. (\ref{eq20}) leads to
\begin{eqnarray}\label{A7}
\begin{aligned}
K_4''=K_7'\cdot A^{(2)} \cdot K_8',
\end{aligned}
\end{eqnarray}
where $K_{7,8}'\in \textrm{exp}[\mathfrak{l}_2(9)]$ and $A^{(2)}\in \textrm{exp}[\mathfrak{a}_2(9)]$.
Then we can let
\begin{eqnarray}\label{A8}
\begin{aligned}
K_7'=\left(
  \begin{array}{ccc}
    V_1 & \textbf{0} & \textbf{0}\\
    \textbf{0} & V_2 & \textbf{0}\\
    \textbf{0} & \textbf{0} & V_2\\
  \end{array}
\right),
K_8'=\left(
  \begin{array}{ccc}
    W_1 & \textbf{0} & \textbf{0}\\
    \textbf{0} & W_2 & \textbf{0}\\
    \textbf{0} & \textbf{0} & W_2\\
  \end{array}
\right),
\end{aligned}
\end{eqnarray}
where $V_{1,2},W_{1,2}\in U(3)$. And $A^{(2)}$ can be written in the following form
\begin{eqnarray}\label{A9}
\begin{aligned}
A^{(2)}=\left(
  \begin{array}{ccc}
    I_3 & \textbf{0} & \textbf{0}\\
    \textbf{0} & T & \textbf{0}\\
    \textbf{0} & \textbf{0} & T^{\dag}\\
  \end{array}
\right),
\end{aligned}
\end{eqnarray}
where $T\in U(3)$ and $T$ is diagonal.
Substituting Eqs. (\ref{A8}) and (\ref{A9}) into Eq. (\ref{A7}), we can get
\begin{eqnarray}\label{A10}
\begin{aligned}
K_4''=K_7\cdot A^{(2)} \cdot K_8'',
\end{aligned}
\end{eqnarray}
where $K_7=I_3\otimes V_2\in \textrm{exp}[\mathfrak{l}_3(9)]$ and
\begin{eqnarray}\label{A11}
\begin{aligned}
K_8''=\left(
  \begin{array}{ccc}
    V_2^{\dag}V_1W_1 & \textbf{0} & \textbf{0}\\
    \textbf{0} & W_2 & \textbf{0}\\
    \textbf{0} & \textbf{0} & W_2
  \end{array}
\right)\in \textrm{exp}[\mathfrak{l}_2(9)].
\end{aligned}
\end{eqnarray}
Finally, following the Cartan decomposition of $\mathfrak{l}_3(9)$ we can obtain
\begin{eqnarray}\label{A12}
\begin{aligned}
K_8''=K_8\cdot A^{(3)}_2 \cdot K_9,
\end{aligned}
\end{eqnarray}
where $A^{(3)}_{2}\in \textrm{exp}[\mathfrak{a}_3(9)]$ and $K_{8,9}\in \textrm{exp}[\mathfrak{l}_3(9)]$.

Substituting Eqs. (\ref{A6}), (\ref{A10}), and (\ref{A12}) into Eq. (\ref{A4}) in turn results in Corollary 1. $\hfill\blacksquare$

Combining the above proof, the rightmost GCX gate in the synthesis of $A$ ($A^{(1)}_{1,2}$) shown in Fig. \ref{Fig.4} (Fig. \ref{Fig.5}) can be absorbed by the block diagonal matrix $K_i'$ in Eq. (\ref{A1}).
Concretely, following the matrix representation of $A$ shown in Eq. (\ref{A3}), we can get
\begin{eqnarray}\label{A13}
\begin{aligned}
A=\left(
  \begin{array}{ccc}
    Z^{01} & \textbf{0} & \textbf{0}\\
    \textbf{0} & I_3 & \textbf{0}\\
    \textbf{0} & \textbf{0} & I_3
  \end{array}
\right)\cdot\left(
  \begin{array}{ccc}
    Z^{01}C_0 & -\textrm{i}Z^{01}S_0 & \textbf{0}\\
    -\textrm{i}S_0 & C_0 & \textbf{0}\\
    \textbf{0} & \textbf{0} & I_3\\
  \end{array}
\right),
\end{aligned}
\end{eqnarray}
where $Z^{01}=\textrm{diag}\{1,-1,1\}$.
Let the left diagonal matrix in the above equation be GCZ and the matrix on the right side be $A'$.
Note that GCZ can be absorbed by $K_2'$ and this does not affect the later deduction.
Furthermore, since
\begin{eqnarray}\label{A14}
\begin{aligned}
\textrm{GCZ}=&R^{01}_y(\frac{\pi}{2})\otimes I_3 \cdot \textrm{GCX}_2^1(1\rightarrow X^{01})\\& \cdot R^{01}_y(-\frac{\pi}{2})\otimes I_3,
\end{aligned}
\end{eqnarray}
combined with the decomposition of $A$ given in Eq. (\ref{eq30}), one can find that
\begin{eqnarray}\label{A15}
\begin{aligned}
A'=&R^{01}_y(\frac{\pi}{2})\otimes I_3
 \cdot R_z^{01}(2\theta_{1}+\theta_{2})\otimes I_3\\&
\cdot \textrm{GCX}_2^1(0\rightarrow X^{01}) \cdot X^{01} \otimes I_3 \\&
\cdot R_z^{01}(\theta_{1}+2\theta_{2})\otimes I_3  \cdot \textrm{GCX}_2^1(2\rightarrow X^{01})\\&
\cdot R_z^{01}(2\theta_{3}-\theta_{1}-\theta_{2})\otimes I_3  \cdot R^{01}_y(-\frac{\pi}{2})\otimes I_3.
\end{aligned}
\end{eqnarray}
Therefore, only two GCX gates are necessary to synthesize $A'$, which means that the rightmost GCX gate in Fig. \ref{Fig.4} can be absorbed.

Likewise, for $A^{(1)}_{i}$ ($i=1,2$), it can be found that
\begin{eqnarray}\label{A16}
\begin{aligned}
A^{(1)}_{i}=&\left(
  \begin{array}{ccc}
    I_3 & \textbf{0} & \textbf{0}\\
    \textbf{0} & Z^{01} & \textbf{0}\\
    \textbf{0} & \textbf{0} & I_3
  \end{array}
\right)\cdot\left(
  \begin{array}{ccc}
    I_3 & \textbf{0} & \textbf{0}\\
    \textbf{0} & Z^{01}C_i & -\textrm{i}Z^{01}S_i\\
    \textbf{0} & -\textrm{i}S_i & C_i\\
  \end{array}
\right).
\end{aligned}
\end{eqnarray}
The left diagonal matrix in the above equation can be decomposed as
\begin{eqnarray}\label{A17}
\begin{aligned}
R^{12}_y(\frac{\pi}{2})\otimes I_3 \cdot \textrm{GCX}_2^1(1\rightarrow X^{12}) \cdot R^{12}_y(-\frac{\pi}{2})\otimes I_3.
\end{aligned}
\end{eqnarray}
Thus, in connection with Eq. (\ref{eq33}) it follows that the rightmost GCX gate in the synthesis of $A^{(1)}_{1}$ ($A^{(1)}_{2}$) shown in Fig. \ref{Fig.5} can be absorbed by $K_1'$ ($K_3'$).

\medskip

\section{Maximally Abelian subalgebra in $\mathbf{\mathfrak{u}(3^\emph{n})}$}\label{Appendix.B}

Note that $\alpha(3^n)$ defined by Eq. (\ref{eq46}) is Abelian as the elements in $\alpha(3^n)$ are all diagonal matrices. The following Lemma is essential for constructing Cartan subalgebras.

\textbf{Lemma.} The $\textrm{span}\{\alpha(3^n)\}$ is maximally Abelian in $\mathfrak{u}(3^n)$.

\textbf{Proof.} Clearly, $\textrm{span}\{\alpha(3)\}$ is maximally Abelian in $\mathfrak{u}(3)$.
Assume that $\textrm{span}\{\alpha(3^{n-1})\}$ is maximally Abelian in $\mathfrak{u}(3^{n-1})$, and we prove below that $\textrm{span}\{\alpha(3^n)\}$ is maximally Abelian in $\mathfrak{u}(3^{n})$.

For any $A\in\mathfrak{u}(3^{n})$, $A$ can be denoted as
\begin{eqnarray}\label{B1}
\begin{aligned}
A=&I_{3}\otimes A_1+\sigma_z^{01}\otimes A_2+\sigma_z^{02}\otimes A_3\\&+\sigma_x^{01}\otimes A_4+\sigma_y^{01}\otimes A_5+\sigma_x^{02}\otimes A_6\\&+\sigma_y^{02}\otimes A_7+\sigma_x^{12}\otimes A_8+\sigma_y^{12}\otimes A_9,
\end{aligned}
\end{eqnarray}
where $A_i\in\mathfrak{u}(3^{n-1})$. Assume $A$ and $\textrm{span}\{\alpha(3^n)\}$ be commutable, then it is sufficient to prove that $A\in\textrm{span}\{\alpha(3^n)\}$.
As $\sigma_z^{01}\otimes I_{3^{n-1}}\in\textrm{span}\{\alpha(3^n)\}$, we have $[A,\sigma_z^{01}\otimes I_{3^{n-1}}]=\textbf{0}$.
From this, we get $A_j=\mathbf{0}, j=4,5,\cdots,9$, that is,
\begin{eqnarray}\label{B2}
\begin{aligned}
A=I_{3}\otimes A_1+\sigma_z^{01}\otimes A_2+\sigma_z^{02}\otimes A_3.
\end{aligned}
\end{eqnarray}

And for any $B\in\textrm{span}\{\alpha(3^{n-1})\}$, then $I_3\otimes B\in \textrm{span}\{\alpha(3^n)\}$.
Thus, we have $[A, I_3\otimes B]=\textbf{0}$, which is equivalent to $[A_k, B]=\textbf{0}$, $k=1,2,3$.
By assumption and the arbitrariness of $B$, it follows that $A_k\in \textrm{span}\{\alpha(3^{n-1})\}$, and then $A\in\textrm{span}\{\alpha(3^n)\}$.
Hence $\textrm{span}\{\alpha(3^n)\}$ is maximally Abelian in $\mathfrak{u}(3^{n})$.
By induction the conclusion holds.$\hfill\blacksquare$

\end{document}